\documentclass[pra,aps,showpacs,groupedaddress,superscriptaddress,twocolumn,toc=flat,biblatex,footinbib,floatfix]{revtex4-1}
\usepackage{natbib}
\usepackage[table]{xcolor}
\usepackage[colorlinks=false]{hyperref}
\usepackage[utf8]{inputenc}
\usepackage{color}
\usepackage{bbm} 
\usepackage[english]{babel}
\usepackage{graphicx}
\usepackage{placeins}
\usepackage{multirow}
\usepackage{csquotes}
\usepackage{physics}
\usepackage{appendix}

\usepackage{amsfonts,amsmath,amssymb,stmaryrd}

\usepackage{graphicx}
\usepackage{subfigure}  % use for side-by-side figures

\usepackage{epsfig}
\usepackage{mathrsfs}
\usepackage{verbatim}
\usepackage{centernot}
\usepackage{ulem}
\usepackage{longtable}
\usepackage{nicefrac}
\usepackage[framemethod=tikz]{mdframed}

\usepackage{tikz}

\newcommand{\singlet}{
  \begin{tikzpicture}[baseline=-0.5ex]
  \begin{scope}[scale=0.8]
    \draw (0,0) ellipse (0.1 and 0.25);
  \end{scope}
  \end{tikzpicture}
}

\newcommand{\holes}{
  \begin{tikzpicture}[baseline=-0.5ex]
  \begin{scope}[scale=0.8]
    \draw (0,-0.1) circle (0.1);
    \draw (0,0.2) circle (0.1);
  \end{scope}
  \end{tikzpicture}
}

\newcommand{\holesigma}{
  \begin{tikzpicture}[baseline=-0.5ex]
  \begin{scope}[scale=0.8]
    \node[label={{$\sigma$}}] at (0,-0.5) {};
    \draw (0,0.2) circle (0.1);
  \end{scope}
  \end{tikzpicture}
}

\newcommand{\sigmahole}{
  \begin{tikzpicture}[baseline=-0.5ex]
  \begin{scope}[scale=0.8]
    \draw (0,-0.1) circle (0.1);
    \node[label={{$\sigma$}}] at (0,-0.2) {};
  \end{scope}
  \end{tikzpicture}
}

\newcommand{\updown}{
  \begin{tikzpicture}[baseline=-0.5ex]
  \begin{scope}[scale=0.8]
    \node[label={{$\downarrow$}}] at (0,-0.7) {};
    \node[label={{$\uparrow$}}] at (0,-0.2) {};
  \end{scope}
  \end{tikzpicture}
}

\newcommand{\downup}{
  \begin{tikzpicture}[baseline=-0.5ex]
  \begin{scope}[scale=0.8]
    \node[label={{$\uparrow$}}] at (0,-0.7) {};
    \node[label={{$\downarrow$}}] at (0,-0.2) {};
  \end{scope}
  \end{tikzpicture}
}

\newcommand{\downsigma}{
  \begin{tikzpicture}[baseline=-0.5ex]
  \begin{scope}[scale=0.8]
    \node[label={{$\sigma$}}] at (0,-0.7) {};
    \node[label={{$\downarrow$}}] at (0,-0.2) {};
  \end{scope}
  \end{tikzpicture}
}

\newcommand{\upsigma}{
  \begin{tikzpicture}[baseline=-0.5ex]
  \begin{scope}[scale=0.8]
    \node[label={{$\sigma$}}] at (0,-0.7) {};
    \node[label={{$\uparrow$}}] at (0,-0.2) {};
  \end{scope}
  \end{tikzpicture}
}

\newcommand{\holeup}{
  \begin{tikzpicture}[baseline=-0.5ex]
  \begin{scope}[scale=0.8]
   \draw (0,0.2) circle (0.1);
    \node[label={{$\uparrow$}}] at (0,-0.7) {};
  \end{scope}
  \end{tikzpicture}
}

\newcommand{\holedown}{
  \begin{tikzpicture}[baseline=-0.5ex]
  \begin{scope}[scale=0.8]
   \draw (0,0.2) circle (0.1);
    \node[label={{$\downarrow$}}] at (0,-0.7) {};
  \end{scope}
  \end{tikzpicture}
}

\usepackage{array}

\usepackage{cancel,ifthen}
\newcommand{\cmnt}[2][NoInPuT]{\ifthenelse{\equal{#1}{NoInPuT}}{}{{\color{red}\sout{#1}}} {\color{blue} #2}}

\usepackage{bm}	% bold in math mode
\renewcommand{\vec}[1]{\bm{#1}}

\LTcapwidth=\textwidth

\begin{document}

\normalem	% changes \emph back to normal after introducing ulem package.

\title{Feshbach resonance in a strongly repulsive bilayer model: \\ a possible scenario for bilayer nickelate superconductors
}

\author{Hannah Lange}
\affiliation{Ludwig-Maximilians-University Munich, Theresienstr. 37, Munich D-80333, Germany}
\affiliation{Max-Planck-Institute for Quantum Optics, Hans-Kopfermann-Str.1, Garching D-85748, Germany}
\affiliation{Munich Center for Quantum Science and Technology, Schellingstr. 4, Munich D-80799, Germany}

\author{Lukas Homeier}
\affiliation{Ludwig-Maximilians-University Munich, Theresienstr. 37, Munich D-80333, Germany}
\affiliation{Munich Center for Quantum Science and Technology, Schellingstr. 4, Munich D-80799, Germany}

\author{Eugene Demler}
\affiliation{Institute for Theoretical Physics, ETH Zurich, 8093 Zürich, Switzerland}

\author{Ulrich~Schollw\"ock}
\affiliation{Ludwig-Maximilians-University Munich, Theresienstr. 37, Munich D-80333, Germany}
\affiliation{Munich Center for Quantum Science and Technology, Schellingstr. 4, Munich D-80799, Germany}

\author{Fabian Grusdt}
\affiliation{Ludwig-Maximilians-University Munich, Theresienstr. 37, Munich D-80333, Germany}
\affiliation{Munich Center for Quantum Science and Technology, Schellingstr. 4, Munich D-80799, Germany}

\author{Annabelle Bohrdt}
\affiliation{Munich Center for Quantum Science and Technology, Schellingstr. 4, Munich D-80799, Germany}
\affiliation{University of Regensburg, Universitätsstr. 31, Regensburg D-93053, Germany}

\pacs{}

\date{\today}

\begin{abstract}
Since the discovery of superconductivity in cuprate materials, the minimal ingredients for high-T$_c$ superconductivity have been an outstanding puzzle. Motivated by the recently discovered nickelate bilayer superconductor La$_3$Ni$_2$O$_7$ under pressure, we study a minimal bilayer model, in which, as in La$_3$Ni$_2$O$_7$, inter- and intralayer magnetic interactions but no interlayer hopping are present: a mixed-dimensional (mixD) $t-J$ model. In the setting of a mixD ladder, we show that the system exhibits a crossover associated with a Feshbach resonance: from a closed-channel dominated regime of tightly bound bosonic pairs of holes to an open-channel dominated regime of spatially more extended Cooper pairs. The crossover can be tuned by varying doping, or by a nearest-neighbor Coulomb repulsion $V$ that we include in our model. Using density matrix renormalization group (DMRG) simulations and analytical descriptions of both regimes, we find that the ground state is a Luther-Emery liquid, competing with a density wave of tetraparton plaquettes at commensurate filling $\delta=0.5$ at large repulsion, and exhibits a pairing dome where binding is facilitated by doping. Our observations can be understood in terms of pairs of correlated spinon-chargon excitations constituting the open channel, which are subject to attractive interactions mediated by the closed channel of tightly bound chargon-chargon pairs. When the closed channel is lowered in energy by doping or tuning $V$, a Feshbach resonance is realized, associated with a dome in the binding energy. Our predictions can be directly tested in state-of-the art quantum simulators, and we argue that the pairing mechanism we describe may be realized in the nickelate bilayer superconductor La$_3$Ni$_2$O$_7$.
\end{abstract}

\maketitle

%%%%%%%%%%%%%%%%%%%%%%%%%%%%%%%%%%%%%%%%%%%%%%%%%%%%%%%%%%%%%

Since the discovery of high-T$_c$ superconductors \cite{Bednorz1986,Lee2006,Scalapino1999,Nozieres:1985} around four decades ago, the search for materials with increasing critical temperatures has lead to the discovery of unconventional superconductivity in a number of compounds, among them copper- and nickel based superconductors \cite{Schilling1993,Li2019}. Very recently, a remarkable critical temperature of $T_c=80\, \mathrm{K}$ was observed in the bilayer nickelate La$_3$Ni$_2$O$_7$ \cite{Sun2023} under pressure, a system with low energy physics that was argued by several groups to be modeled by a bilayer $t-J$ model with weak hopping strength but strong antiferromagnetic Heisenberg couplings between the layers \cite{Lu2023,Oh2023type,Qu2023,Wu2023}. 

\begin{figure}[t]
\centering
\includegraphics[width=0.45\textwidth]{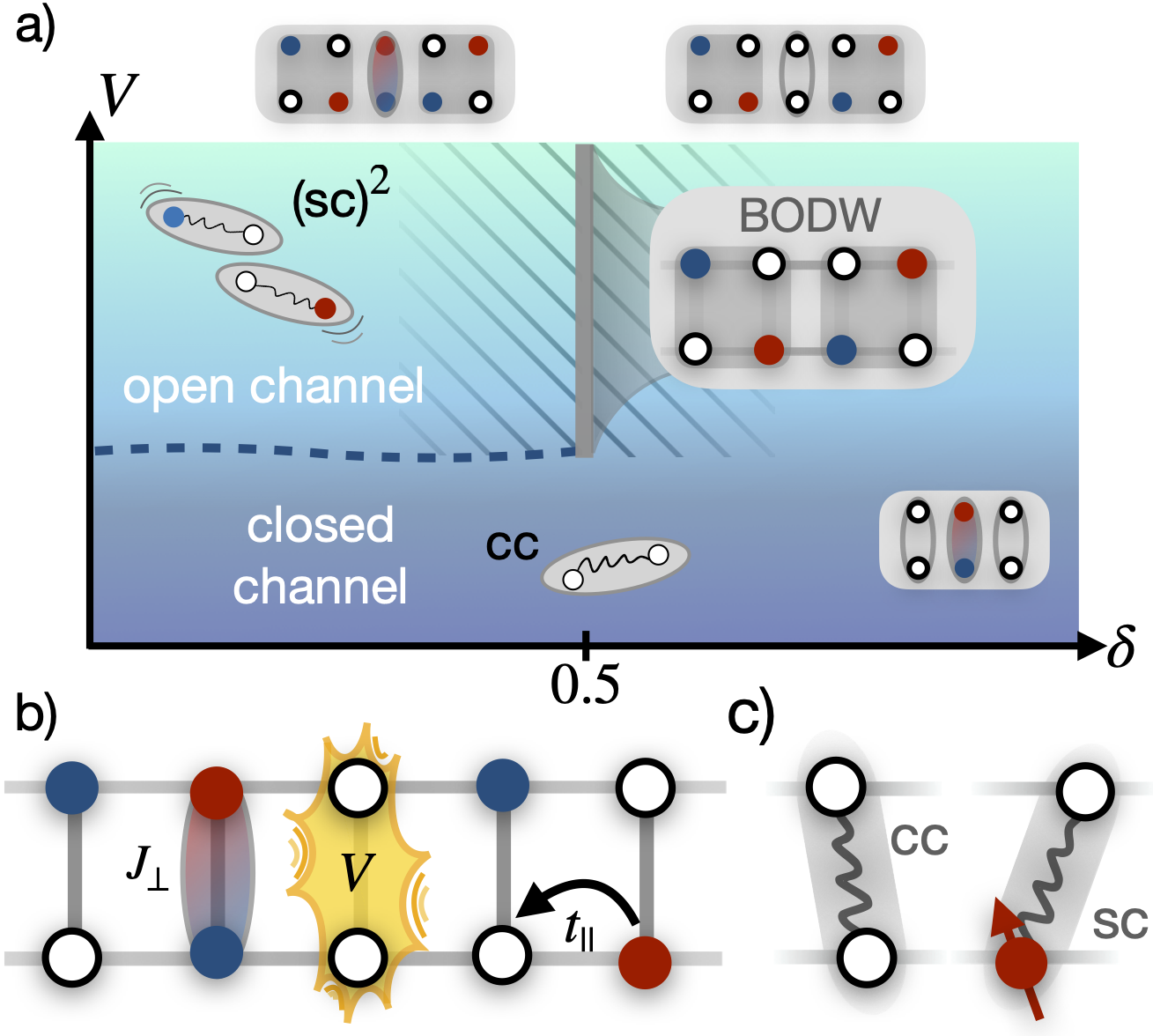}
\caption{a) Sketch of the phase diagram of the mixed dimensional ladder with repulsive interactions $V$ shown in b). We predict a crossover associated with a Feshbach resonance tuned by doping or $V$, which can be understood in terms of the emergent mesonic charge carriers described in the text and sketched in c). For large values of $V$, the tightly bound chargon-chargon (cc) pairs, previously reported for mixD ladders without repulsion \cite{Bohrdt2022,Hirthe2023}, become spatially more extended and can be understood in terms of spinon-chargon (sc) pairs that experience an attraction arising from a Feshbach resonance \cite{lange2023} and leading to the formation of $(\mathrm{sc})^2$ Cooper pairs. Around a hole doping of $\delta=50\%$ and for values of $V$ where the low-doping side is in the open channel dominated $(\mathrm{sc})^2$ regime, the bond-ordered density wave (BODW, illustrated in inset) emerges, which is a correlated spinon-chargon phase with spin and leg index order and a charge gap.} 
\label{fig:1}    
\end{figure}

Motivated by this minimal working example of unconventional superconductivity, understanding the pairing mechanism of mixed-dimensional (mixD) Fermi-Hubbard or $t-J$ bilayers, or ladders, is an important step towards a microscopic theory of pairing \cite{Grusdt2018_2,Bohrdt2021, Bohrdt2022}. MixD systems, featuring magnetic superexchange in $d$ dimensions but hopping only in $d-1$ dimensions, see Fig.~\ref{fig:1}b, have been shown to host the following emergent structures upon doping the ground state at half-filling, consisting of singlets on each rung of the ladder: $(i)$ When doped with a single hole, the system can be understood as a mesonic spinon-chargon (sc) bound state of two partons, a charge excitation (\textit{chargon}) and a spin excitation (\textit{spinon}), carrying the respective quantum numbers and being connected by a linear, \textit{string}-like confinement potential \cite{Beran1996,Trugman1988,Brinkman1970,Laughlin1997,Senthil2003, Grusdt2018, Grusdt2019,Chiu2019,Bohrdt2021, Bohrdt2022}, see Fig.~\ref{fig:1}c (right). Hereby, the string has its origin in the disruption of the ground-state singlet order when the charge moves through the system, see Appendix \ref{appendix:mixDbinding}. $(ii)$ Owing to a similar mechanism, two holes form a tightly bound state of two chargons in the mixD setting without repulsion. Here, strong binding energies emerge due to the fact that it is favorable for the chargons to move through the system coherently, since a second chargon can restore the distorted spin order by following the first chargon \cite{Bohrdt2022} (\textit{chargon-chargon}, cc pairs), see Fig. \ref{fig:1}c (left) and Appendix \ref{appendix:mixDbinding}. The formation of the tightly bound chargon-chargon pairs, with a large binding energy, has allowed their direct observation in ultracold atom experiments by Hirthe~et~al.~\cite{Hirthe2023}. 

In this article we show that the mixD setting allows to tune through a crossover associated with a Feshbach resonance, which enables strong pairing despite the presence of dominant Coulomb repulsion \cite{lange2023}. On one side of the crossover we find a BEC-like, i.e. closed-channel dominated, regime of chargon-chargon pairs; on the other side, i.e. the BCS side, of the resonance an open-channel dominated regime of strongly correlated and spatially more extended spinon-chargon pairs $(\mathrm{sc})^2$ is realized, see Fig.~\ref{fig:1}. This allows us to study binding in the strongly correlated mixD electron systems from the perspective of a BEC to BCS crossover.

In atomic systems, where the BEC to BCS crossover has initially been explored using ultracold atom experiments \cite{Drechsler1992, Greiner:2003, Zwierlein2003,Jochim2003,Bloch2008,Gross2017}, the transition from the open-channel dominated Luther-Emery liquid of paired fermions to the closed-channel dominated regime of bosonic molecules \cite{Tokatly2004,Recati2005,Fuchs2004} occurs via a Feshbach resonance that couples open and closed channels \cite{Citro2005}. Hereby, attractive interactions in the open channel are mediated by processes that couple to the high-energy closed channel. We argue that a similar scenario is naturally realized in mixD doped antiferromagnets -- and, by extension, in the bilayer nickelate superconductor La$_3$Ni$_2$O$_7$ -- albeit the underlying constituents are emergent spin-charge composits (namely the sc and cc excitations introduced above) instead of the underlying microscopic fermions of the model. Thereby an instance of mesonic Feshbach resonance \cite{Homeier2023,lange2023} is realized, related to similar ideas how emergent Feshbach resonances can induce unconventional pairing interactions in strongly correlated electron systems \cite{Crepel2023}.

Starting from the closed-channel dominated side with cc constituents and at low doping, we increase the Coulomb repulsion $V$ to tune through the crossover and into the open-channel dominated regime of spatially more extended $\mathrm{sc}$ pairs. The latter remain bound into $(\mathrm{sc})^2$ Cooper-like pairs through the pairing mechanism realized in Feshbach resonances \cite{Feshbach1958}: recombinations of two sc pairs into a closed-channel cc state mediates an attractive interaction among sc's. This scenario allows to overcome strongly repulsive Coulomb interactions, and we find large binding energies on the open-channel dominated side of the resonance with a peak at $30\%$ doping when $V$ is the dominant energy scale \cite{lange2023}. 

Alternatively, we can tune across the Feshbach resonance at large $V$ by varying doping $\delta$. Beyond $\delta > 50\%$ closed-channel cc states proliferate, which boosts the induced attraction among sc constituents and leads to the formation of tightly bound $(\mathrm{sc})^2$-type Cooper pairs, with tetraparton character and residing on plaquettes of the ladder, as illustrated in Fig.~\ref{fig:1}a. Furthermore, right at the commensurate hole doping $\delta = 50\%$, the strongly correlated nature of the Cooper pairs leads to a competition with a bond-ordered density wave (BODW) featuring both charge and spin gaps in the one-dimensional ladder geometry.

The remaining part of the paper is organized as follows: In the first section, we introduce the mixD $t-J$ ladder model supplemented by repulsive interactions~$V$ (mixD+V, see Fig. \ref{fig:1}b). We explain how it can be realized experimentally with ultracold atoms, and how it relates to recently discovered high-$T_c$ superconductivity in bilayer nickelates. Second, we analyze the cc (closed channel) and sc (open channel) limits analytically and derive the induced $s$-wave attraction among the latter. Then, we present numerical results from DMRG simulations \cite{White1992,Schollwoeck2011} on the crossover between the cc and the sc regimes and find a good agreement with the effective cc and sc descriptions. Lastly, we show that the system forms a Luther-Emery liquid everywhere away from $\delta=50\%$ and describe the BODW state found at the commensurate filling $\delta=50\%$ for strong repulsion.

\FloatBarrier
\section{Mixed dimensional system with Coulomb repulsion}
\subsection{Effective nickelate bilayer model}
The mixD+V model that we investigate is shown schematically in Fig. \ref{fig:1}b. It consists of nearest neighbor hopping along the legs of the ladder with amplitude $t_\parallel$, superexchange interactions $J_\perp$ and repulsive Coulomb interactions $V$ for neighbors on rungs, i.e.
\begin{align}
    \hat{\mathcal{H}}&=-t_{\parallel}\hat{\mathcal{P}}\sum_{j}\sum_{\mu, \sigma}\left( \hat{c}_{j+1\mu\sigma}^\dagger \hat{c}_{j\mu\sigma} + \mathrm{h.c.}\right)\hat{\mathcal{P}}\notag \\
    &+ J_{\perp}\sum_{j }\left(\hat{\vec{S}}_{j0}\cdot \hat{\vec{S}}_{j1} -\frac{1}{4}\hat{n}_{j0}\hat{n}_{j1}\right) +V \sum_{j }\hat{n}_{j0}^h \hat{n}_{j1}^h\, ,
        \label{eq:mixD_Hamiltonian}
\end{align}
where $\hat{\mathcal{P}}$ is the Gutzwiller projector that projects onto the subspace with maximum single occupancy per site. Spin and (hole) density operators at site $i$ in layer $\mu=0,1$ are denoted by $\hat{\vec{S}}_{i\mu}$ and $\hat{n}_{i\mu}=\hat{n}_{i\mu\uparrow}+\hat{n}_{i\mu\downarrow}$ ($\hat{n}_{i\mu}^h=1-\hat{n}_{i\mu}$).

The ground state of the model without hole doping, $\delta=0$, consists of one singlet on each rung of the ladder, with energy $E_0^{\delta=0}=-L_x J_\perp$. Upon doping, the physics of this model is determined by a competition between kinetic, magnetic and Coulomb contributions. The emergent constituents can be most easily understood in the tight-binding regime $0 \approx t_\parallel\ll J_\perp$ without Coulomb repulsion, $V=0$: In this limit, it is favorable for two holes in opposite legs to sit on neighboring sites to reduce the number of rungs with distorted rung-singlet spin configuration. For $t_\parallel \gtrsim J_\perp$, the chargon-chargon pairs develop a spatial structure. However, it is still advantageous for holes to move through the system coherently, since the second hole can retrace the distorted spin background of the first hole, see Appendix~\ref{appendix:mixDbinding}, yielding tightly bound pairs of holes that are close to each other in real space \cite{Hirthe2023,Bohrdt2022}. 

When the repulsive interaction $V$ reaches a critical value $V_c>0$, it is energetically favorable to place at maximum one spin and one hole per rung, i.e. to form spinon-chargon pairs. The respective regimes are dominated by (i) bosonic \textit{chargon-chargon} (cc) pairs and (ii) fermionic \textit{spinon-chargon} (sc) pairs, which themselves pair up to form (sc)$^2$-type Cooper pairs, see Fig. \ref{fig:1}a. We will refer to the two regimes as closed- and open channel dominated, respectively. Note that each of the emergent mesonic charge carriers, cc's and sc's, can be assumed point-like for $t_\parallel \ll J_\perp$, i.e. with constituents on the same rung, but develop an internal spatial structure for $t_\parallel\gtrsim J_\perp$ \cite{Bohrdt2022}.

The mixD+V model is closely related to the recently discovered, pressure-induced nickelate superconductor La$_3$Ni$_2$O$_7$ \cite{Sun2023,Lu2023,Oh2023type,Qu2023,Wu2023,Gu2023,Luo2023,Lu2023_}. Density functional theory calculations for this material have shown that the low energy physics is determined by the $d_{x^2-y^2} $ and $d_{z^2}$ orbitals \cite{Sun2023,zhang2023,christiansson2023,Shilenko}. The $d_{x^2-y^2}$ orbitals form an effective intralayer $t-J$ model, whereas the $d_{z^2}$ orbitals are localized with interlayer antiferromagnetic (AFM) superexchange through the apical $p$-orbital of the intermediate oxygen layer. This interlayer coupling is enhanced under pressure, when the angles between Ni in opposite layers and O changes. Both orbitals interact with each other via ferromagnetic Hund's couplings. In the limit of large Hund's coupling, the spins of $d_{x^2-y^2} $ and $d_{z^2} $ align, giving rise to an effective AFM interaction $J_\perp$ between the layers \cite{Lu2023,Oh2023type,Qu2023}. 

In contrast to AFM interactions that originate only from superexchange, the interaction mediated via Hund's coupling corresponds to a vanishingly small interlayer hopping $t_\perp$. We further argue that at low doping, when Coulomb interactions are not yet fully screened, the nearest-neighbor repulsion $V$ between the layers can potentially play an important role.

The mixD+V model in Eq.~\eqref{eq:mixD_Hamiltonian} provides a minimal model that can potentially capture some of the essential physics realized in the bilayer nickelates. The model itself can straightforwardly be extended from the numerically easily accessible two-leg ladders to a two-dimensional mixD bilayer setup, as discussed originally in the context of ultracold atom experiments but without the $V$-term in \cite{Bohrdt2021}. The nature of some phases we discuss below is expected to change when going from two-leg ladders to a full-blown bilayer geometry; however, the emergent sc and cc constituents as well as the Feshbach resonance endowing them with attractive interactions, even when the tightly bound cc state is not the ground state, are robust features that we expect to underlie the rich physics of mixD bilayer settings involving extended two-dimensional layers - and potentially high-$T_c$ superconductivity observed in bilayer nickelates.

%-----------------------------------------
\begin{figure}[t]
\centering
\includegraphics[width=0.5\textwidth]{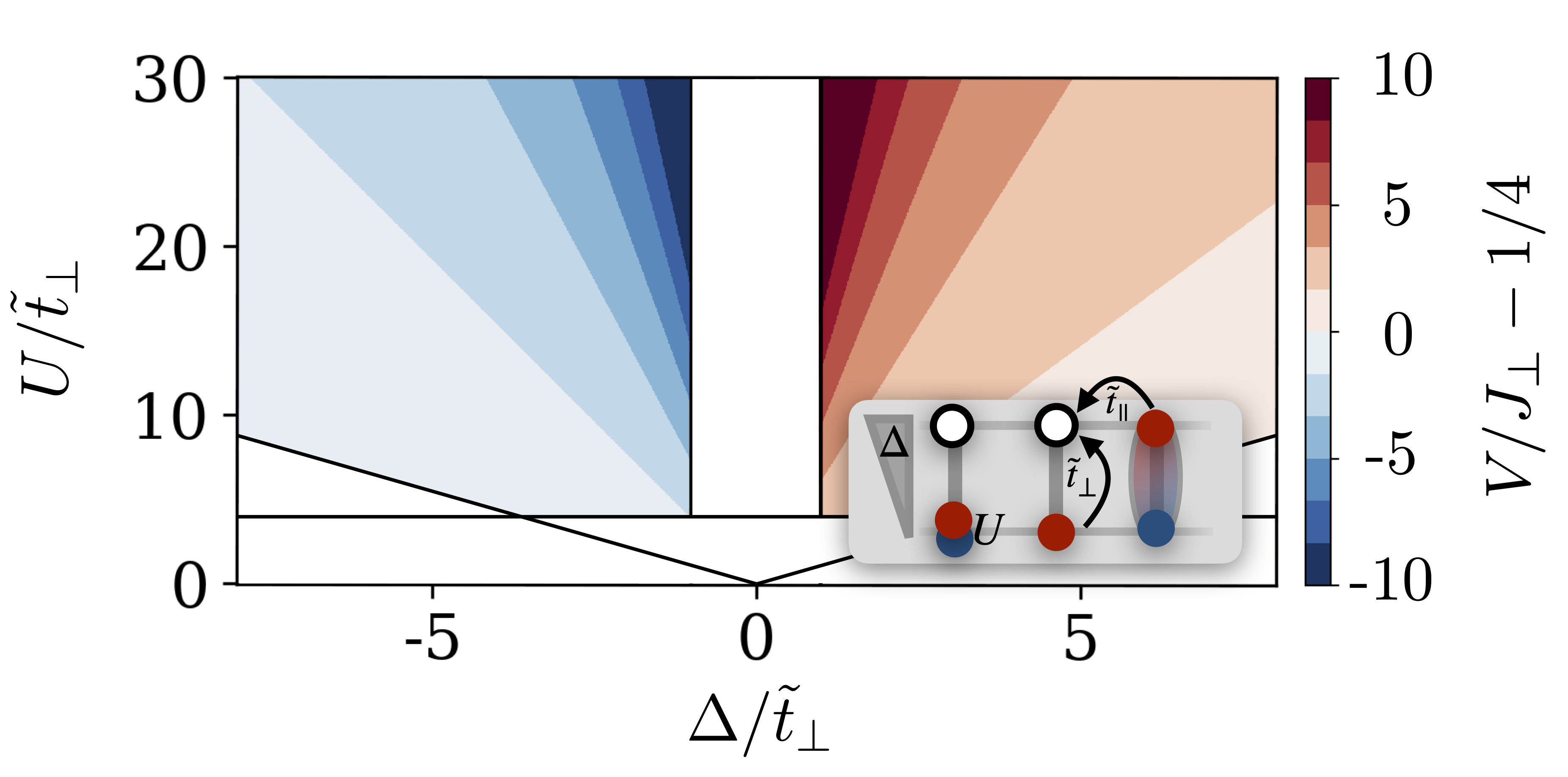}
\caption{Realizing the mixD+V model with ultracold atoms. We predict rung-nearest neighbor repulsion (red) that can be experimentally realized by doping the energetically higher (lower) leg of a tilted Heisenberg ladder with holes (doublons), see inset. In the opposite case, realized for $\Delta < 0$, the energetically higher (lower) leg is doped with doublons (holes) and nearest-neighbor attraction (blue) is realized, see  Eq. \eqref{eq:V}. We plot only regimes where perturbative treatments of $\tilde{t}_\perp$ are reasonable.} 
\label{fig:2}    
\end{figure}
%-------------------------------------------

\subsection{Experimental realization in ultracold atoms}
The mixD$+$V model \eqref{eq:mixD_Hamiltonian} is particularly intriguing because it can be realized in ultracold atom experiments by a modification of the mixD setup by Hirthe et al. \cite{Hirthe2023}. In this cold atom experiment, a potential offset $\Delta$ between the upper and lower leg of a Fermi-Hubbard ladder was applied, with large on-site repulsion $U\gg\Tilde{t}_\parallel, \Tilde{t}_\perp$ and $U>\Delta \gg\Tilde{t}_\perp$, where $\Tilde{t}_\parallel, \Tilde{t}_\perp$ denote the hopping amplitudes and $U$ the on-site repulsion in the Fermi-Hubbard setup. This yields a suppressed tunneling between the chains to effectively $t_\perp=0$, while realizing super-exchange of strength $J_\perp = 4 \tilde{t}_\perp^2 U / (U^2 - \Delta^2)$. 

In Appendix \ref{appendix:Experiment} we show that by replacing doped holes by doublons in the lower leg of the ladder, we can effectively supplement the mixD model with repulsive interactions, see inset of Fig. \ref{fig:2}. This gives rise to \emph{virtual} hopping processes of doublons from the lower leg onto holes in the upper leg, contributing an energy shift $2\frac{\Tilde{t}_\perp^2}{U-\Delta}$; and of doublons in the lower leg onto spins in the upper leg (or of a hole in the upper leg onto a spin in the lower leg) with $-\frac{\Tilde{t}_\perp^2}{\Delta}$. In total, we obtain nearest-neighbor interactions between dopants on the rung with strength $V$ given by 
\begin{align}
    V - \frac{J_\perp}{4}=\Tilde{t}_\perp^2\left(\frac{2}{\Delta}+\frac{U+2\Delta}{U^2-\Delta^2}\right)\,.
    \label{eq:V}
\end{align}

This interaction is positive, i.e. repulsive, for doublon doping in the lower leg and hole doping in the upper leg; vice versa, for doublon doping in the upper leg and hole doping in the lower leg, the resulting interaction obtained by replacing $\Delta$ with $-\Delta$ in Eq.~\eqref{eq:V} is attractive when $U > |\Delta|$. As we show in Fig. \ref{fig:2}, the achievable repulsion and attraction strengths $V$ can reach sizable values in units of $J_\perp$, for feasible parameters $\Tilde{t}_\perp$, $U$ and $\Delta$ in regimes where perturbation theory in $\tilde{t}_\perp$ is reasonable. Notably, we do not require $V/t_\perp$ to be large, which could only be realized in a regime where perturbation theory breaks down.

\subsection{Effective sc and cc descriptions \label{sec:effectiveModels}}
In order to gain a qualitative understanding of the physics in the mixD+V model in both weak and strong repulsion limits, we derive effective low-energy Hamiltonians in terms of cc's $\hat{b}_i^{\dagger}$ in the low $V$ regime and sc's $\hat{f}_i^{(\dagger)}$ in the large $V$ limit. This is done by performing Schrieffer-Wolff transformations of Eq. \eqref{eq:mixD_Hamiltonian} in the respective low energy subspaces and restricting to configurations that only involve point-like cc's and sc's that sit on the same rung $i\in\{0,\dots,L_x\}$. By doing so, we obtain explicit expressions for the effective repulsion between cc's in the low $V$ regime and the attractive interaction between sc's in the strong $V$ regime.

\subsubsection{The closed channel chargon-chargon regime}
In the chargon-chargon (weak $V$) regime, the low-energy subspace is given by holes on the same rung, whereas the high-energy subspace contains spinon-chargon configurations.
The effective Hamiltonian we obtain in this regime,
\begin{align}
\hat{\mathcal{H}}^{cc}_\mathrm{eff} &= -2\frac{t_\parallel^2}{J_\perp-V}\sum_{j }\hat{\mathcal{P}}_b \left(\hat{b}_{j+1}^\dagger \hat{b}_j +\mathrm{h.c.}\right)\hat{\mathcal{P}}_b\notag \\
&+4\frac{t_\parallel^2}{J_\perp-V}\sum_{j } \hat{b}_{j+1}^\dagger\hat{b}_{j+1} \hat{b}_j^\dagger \hat{b}_j
    -\epsilon_0^{cc} \sum_{j }\hat{b}_j^\dagger \hat{b}_j\, ,
    \label{eq:H_eff_cc}
\end{align}
describes hard-core, bosonic cc's with a chemical potential $\epsilon_0^{cc}=z2\frac{t_\parallel^2}{J_\perp-V}+J_\perp -V$; $z$ denotes the coordination number of the lattice, i.e. $z=2$ for the ladder and $z=4$ in the  two-dimensional bilayer geometry. The second term corresponds to a nearest-neighbor repulsion along the legs / in plane. A detailed derivation can be found in Appendix~\ref{appendix:Derivation_cc_Model}.

\subsubsection{The open channel spinon-chargon regime}
In the spinon-chargon regime (strong $V$ and doping $\delta \leq 50\%$) all configurations involving two holes on the same rung can be integrated out. The effective Hamiltonian becomes \cite{lange2023}
\begin{align}           
\mathcal{\hat{H}}^{sc}_\mathrm{eff}&= \frac{t_\parallel}{2}\sum_{j} \sum_{\sigma,\mu} \mathcal{\hat{P}}_f\left(\hat{f}^\dagger_{j+1\mu\sigma}\hat{f}_{j\mu\sigma}+\mathrm{h.c.}\right)\mathcal{\hat{P}}_f \notag \\
&+\epsilon^{sc}_0\sum_{j\mu} \hat{n}^f_{i\mu}
+\frac{t_\parallel^2}{J_\perp}\frac{3}{2}\sum_{j}\sum_{\mu \mu^\prime}\hat{n}^f_{j+1\mu}\hat{n}^f_{j\mu^\prime}
%+V\sum_j \hat{n}^f_{j\mu}\hat{n}^f_{j\bar{\mu}}
\notag \\
& -4t_\parallel^2\sum_{j}\left(
-\hat{\vec{J}}_{j+1}\cdot \hat{\vec{J}}_j+\frac{1}{4}
\right)\left[\frac{\mathcal{\hat{P}}^S_j}{V-J_\perp }+\frac{\mathcal{\hat{P}}^T_j}{V }\right].
\label{eq:H_eff_sc}
\end{align}
with $\epsilon_0^{sc}= J_\perp -\frac{t_\parallel^2}{J_\perp}\frac{3}{2}$ as well as the singlet and triplet projectors $\mathcal{\hat{P}}_S= -\hat{\vec{S}}_{i}\cdot \hat{\vec{S}}_{j}+\frac{1}{4}\hat{n}^f_{i}\hat{n}^f_{j}$ and $\mathcal{\hat{P}}_T= \hat{\vec{S}}_{i}\cdot \hat{\vec{S}}_{j}+\frac{3}{4}\hat{n}^f_{i}\hat{n}^f_{j}$,
where we have defined the spinon-chargon density operators $\hat{n}^f_{i}=\hat{f}_{i\mu\sigma}^\dagger \hat{f}_{i\mu\sigma}$, the
spin operators 
\begin{align}
\hat{\vec{S}}_{i}=\frac{1}{2}\sum_{\mu}\sum_{\sigma\sigma^\prime}\hat{f}_{i\mu\sigma}^\dagger\vec{\sigma}_{\sigma\sigma^\prime}\hat{f}_{i\mu\sigma^\prime}
\end{align}
and rung isospin operators 
\begin{align}
    \hat{\vec{J}}_{i}=\frac{1}{2}\sum_{\sigma} \sum_{\mu\mu^\prime}\hat{f}_{i\mu\sigma}^\dagger\vec{\sigma}_{\mu\mu^\prime}\hat{f}_{i\mu^\prime\sigma}\, .
    \label{eq:J_operators}
\end{align}

Eq. \eqref{eq:H_eff_sc} describes hard-core, fermionic sc's with a chemical potential $\epsilon^{sc}_0$, experiencing competing repulsion $\propto \frac{t_\parallel^2}{J_\perp}$ and attraction $\propto -\frac{t_\parallel^2}{V-J_\perp}$ and $\propto -\frac{t_\parallel^2}{V}$, for singlet-triplet recombination processes to the chargon-chargon channel and back, respectively. We would like to emphasize that the attractive terms arise from second-order recombination processes of sc's to a spin singlet and a cc and back to the sc channel, i.e. the attraction is mediated by virtual coupling processes to the tightly bound, high energy chargon-chargon channel \cite{lange2023}. This phenomenology is in analogy to Feshbach resonances, where attraction in the open channel is induced by the proximity of the closed channel in parameter space. As will be discussed later in Sec. \ref{sec:binding}, the sc model predicts dominant attraction under certain conditions, yielding effective finite binding energies between the constituents. In the perturbative regime assumed in Eq. \eqref{eq:H_eff_sc}, the resonance itself occurs at $V\to J_\perp$, where the attractive interaction diverges and cc’s proliferate.

Moreover, we point out that Eq. \eqref{eq:H_eff_sc} is $SU(2)$ symmetric in the spin and isospin sector. While the $SU(2)$ spin symmetry is already present in the original model \eqref{eq:mixD_Hamiltonian}, the isospin $SU(2)$ is an artifact of the perturbation theory and breaks down if higher orders in $t_\parallel/(V-J_\perp)$ are considered. We will show in Sec. \ref{sec:JJSS} that the isospin $SU(2)$ symmetry is nonetheless approximately present in the numerical results for large $V$, and exhibits a strong doping dependence.

%-----------------------------------------
\begin{figure}[t]
\centering
\includegraphics[width=0.45\textwidth]{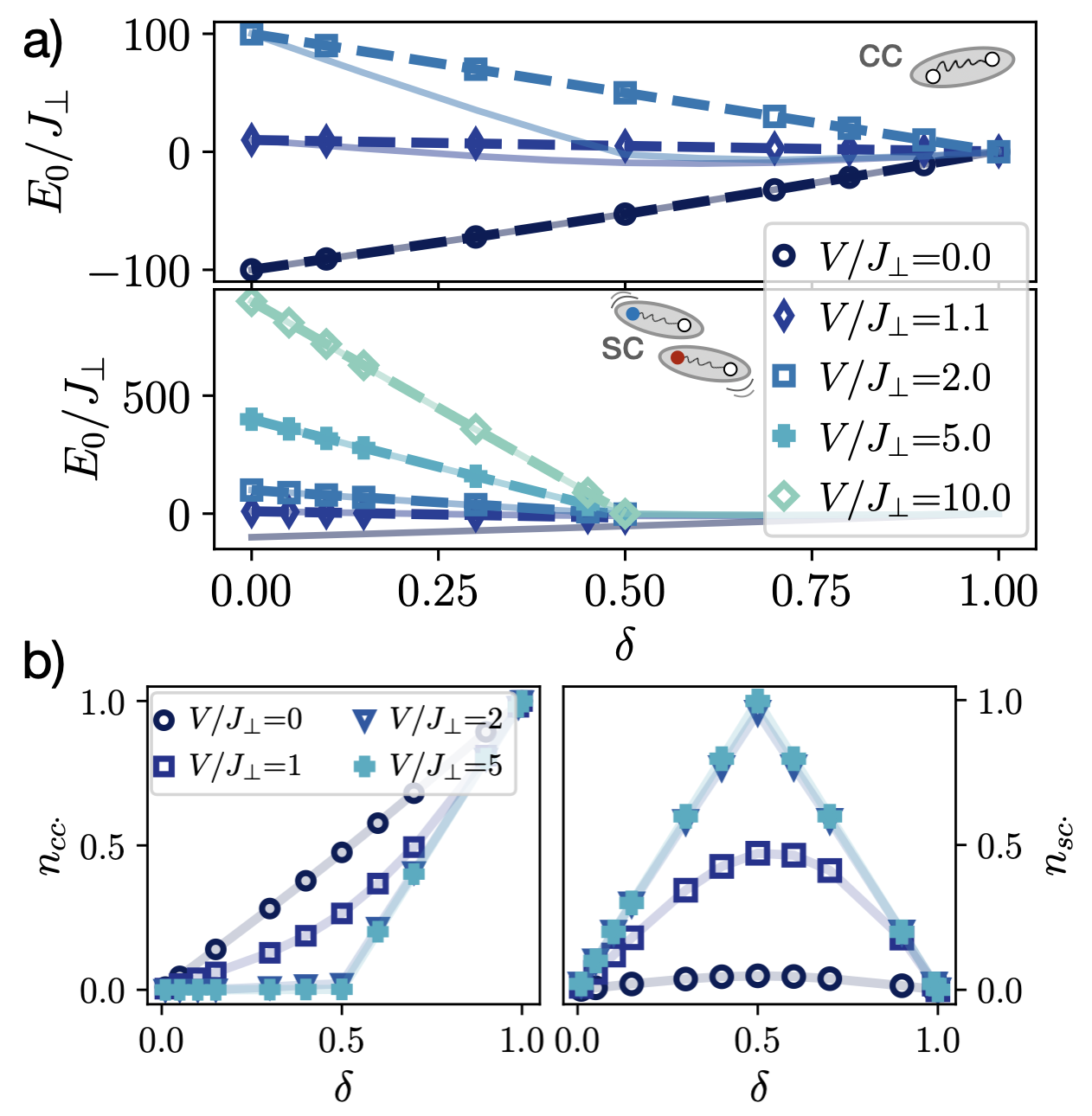}
\caption{a) Ground state energies of the mixD+V ladder (light lines) and the effective cc \eqref{eq:H_eff_cc} (top) and sc \eqref{eq:H_eff_sc} (bottom) model in the tight binding regime $t_\parallel/J_\perp=0.1\ll1$ (data points connected with dashed lines). The derivation of the respective effective models assume point-like charge carriers, with densities $n_{cc}$ and $n_{sc}$ shown in b), calculated by DMRG of the mixD+V model (data points connected with light lines). In a) we find good agreement of the DMRG in the respective regimes: $V/J_\perp < 1$ closed channel (cc) dominated,  $V/J_\perp > 1$ open channel (sc) dominated.} 
\label{fig:7}    
\end{figure}
%-------------------------------------------

Figure \ref{fig:7}a shows the ground state DMRG energies $E_0$ of the effective low-energy descriptions \eqref{eq:H_eff_sc} and \eqref{eq:H_eff_cc} (dashed lines) for small $t_\parallel/J_\perp=0.1\ll1$, compared to the results for the full mixD$+$V model~\eqref{eq:mixD_Hamiltonian} (light, solid lines). The ground state energies are in very good agreement with the full mixD$+$V model in the respective regimes of $V$. The different regimes can be distinguished by analyzing the densities of point-like cc's and sc's, $n_{cc}$ and $n_{sc}$ (calculated from the full mixD+V model). Which one dominates in the weak and strong $V$ regimes, see Fig. \ref{fig:7}b, is in agreement with the respective low energy subspaces. Here, $\hat{n}_{cc} = \frac{2}{L_x} \sum_i^{L_x} \hat{n}^h_{i,\mu=0}\hat{n}^h_{i,\mu=1}$ and $\hat{n}_{sc} = \frac{1}{L_x}\sum_i^{L_x}\sum_{\mu} \hat{n}^h_{i,\mu}(1-\hat{n}^h_{i,\bar{\mu}})$.

For $V=0$, the number of cc's grows linearly with the hole density $\delta$, and $n_{sc}$ is suppressed to very small values. In contrast, for larger repulsion strengths the formation of cc's is suppressed in the intermediate doping regime, with $n_{cc}\approx 0$ below $\delta\leq 0.5$ for $V\geq 2 J_\perp$. In this strongly repulsive regime, we find that the number of sc's increases linearly to a value of $n_{sc}=1$ at $\delta=0.5$, i.e. holes will avoid to sit on the same rung if they can. Above $\delta=0.5$, there are more holes than particles in the system and hence $n_{cc}$ increases to unity ($n_{sc}$ decreases to zero).

Note that we calculate the density of point-like sc (cc) pairs here, as assumed in the derivations. In principle, in particular for larger $t_\parallel$, the distance between spinons and chargons (chargons and chargons) can be larger and spatially overlapping mesons are possible as well (see Fig.~\ref{fig:hole_distance} in Appendix \ref{AppdxHoleDistance}).

\subsubsection{A regime of correlated spinon-chargon pairs: The bond-ordered density wave (BODW) at $\delta=0.5$ \label{sec:BODW}}
For doping $\delta=0.5$, Eq. \eqref{eq:H_eff_sc} takes the form
\begin{align}
    \hat{\mathcal{H}}_\mathrm{eff}^{sc}(\delta=0.5)=-4\frac{t_\parallel^2}{V} &\sum_{\langle ij\rangle}\left(-\vec{\hat{J}}_{i}\cdot \vec{\hat{J}}_{j}+\frac{1}{4} \right)\left(1+\frac{J_\perp}{V}\hat{\mathcal{P}}_S\right),
    \label{eq:Heff_sc_Vlarge_delta0.5}
\end{align}
which we derive in Appendix \ref{appendix:BOWD}. Individually, the two factors under the sum would favor a Heisenberg AFM order of isospins $\vec{\hat{J}}$ and spins $\hat{\mathbf{S}}$, respectively. However, as we have shown in \cite{lange2023}, the product of both terms leads to aground state which is a correlated valence-bond crystal of spins $\vec{\hat{S}}$ and isospins $\vec{\hat{J}}$, with an alternating pattern of singlets (no singlets) on bonds $\langle 2j,2j+1 \rangle $ ($\langle 2j+1,2j+2 \rangle $) for $\vec{\hat{S}}$ and $\vec{\hat{J}}$ sectors. In Appendix \ref{appendix:BOWD} we present a variational argument how the VBS phase emerges. 

Below, we will show that indications for this \textit{bond-ordered density wave (BODW)}, illustrated in the inset of Fig. \ref{fig:1}, can be observed in the numerics, robust to finite size scaling. We will further show that away from the commensurate filling $\delta=50\%$, the numerical results can be understood in terms of domain wall excitations on top of the BODW.

\section{Feshbach resonance and crossover}
In the remaining part of the paper we will present our numerical analysis of the Feshbach resonance and the associated crossover between the open- and closed channel dominated regimes. Thereby we reveal the underlying pairing mechanism between doped holes, and how it changes with increasing Coulomb repulsion $V$ and doping. All our numerical results have been obtained using the DMRG package SyTen \cite{syten1,syten2}. 

The key finding is that the binding energy is positive in the entire doping regime even if strong repulsive interactions dominate. Moreover, the value $c=1$ of the central charge and the Fermi momentum $k_F=\pi \delta/2$ we find in the gapless regimes, i.e. away from the commensurate filling $\delta=50\%$, indicates that the system realizes a Luther Emery liquid constituted by the emergent (sc)$^2$ or cc's.
Therefore, we argue that by tuning the repulsion~$V$ and doping, the character of the ground state changes from open channel dominated (BCS side of the Feshbach resonance) to closed channel dominated (BEC side of the Feshbach resonance), see Figure~\ref{fig:1}. 

%-----------------------------------------
\begin{figure}[t]
\centering
\includegraphics[width=0.45\textwidth]{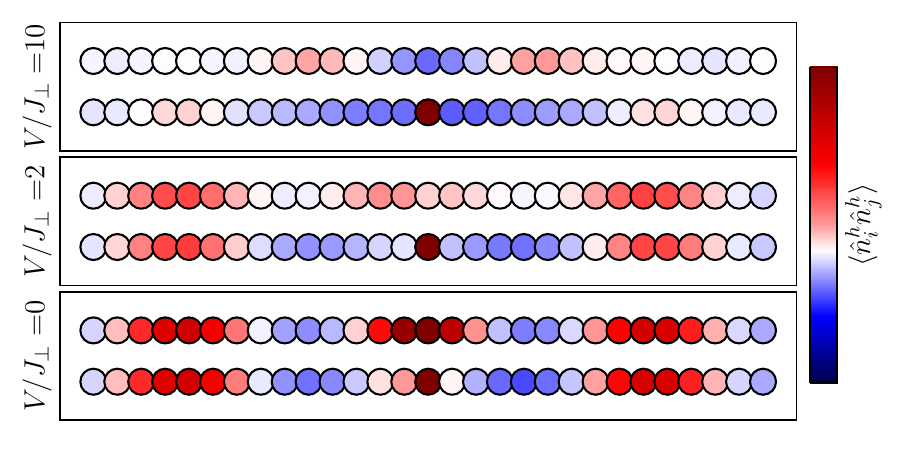}
\caption{Hole density correlation function $\langle \hat{n}^h_i\hat{n}^h_j\rangle$ for a system of length $L_x=200$ and hole doping $\delta=0.9$, site $i$ located in the middle of the system ($x_i=100$) and $t_\parallel/J_\perp=1.0$. The lower panel shows the results for $V/J_\perp=0$, with increasing values of $V$ for the middle and top panels.} 
\label{fig:3}    
\end{figure}
%-------------------------------------------

\subsection{Hole density correlation function}
In order to further investigate the emergent excitations in the low and high Coulomb repulsion regime, we calculate the hole density correlation function $\langle \hat{n}^h_i\hat{n}_j^h\rangle$ with $i=L_x/2$, as exemplary shown for a hole doping $\delta=0.9$ in Fig. \ref{fig:3}.

For low repulsion strengths $V\leq 2 J_\perp$, one can observe a density wave pattern of chargon-chargon pairs with a wavelength $\lambda\approx 10$, with alternating enhanced (red) and suppressed (blue) hole density simultaneously on both chains of the ladder, in agreement with the high density of cc's found in Fig. \ref{fig:7}. 

In contrast, the upper panel of Fig. \ref{fig:3} shows that for large repulsion, the probability of finding a hole in the upper chain at distance $\vert j-i\vert $ is suppressed if there is a hole in the lower chain, and vice versa, since holes on the same rung get an energy penalty on the order of $V$, resulting in a density wave of spinon-chargon pairs with finite-range correlations if $\delta\neq 0.5$, see Fig. \ref{fig:3} top.

\subsection{Binding energies \label{sec:binding}}
\begin{figure}[t]
\centering
\includegraphics[width=0.49\textwidth]{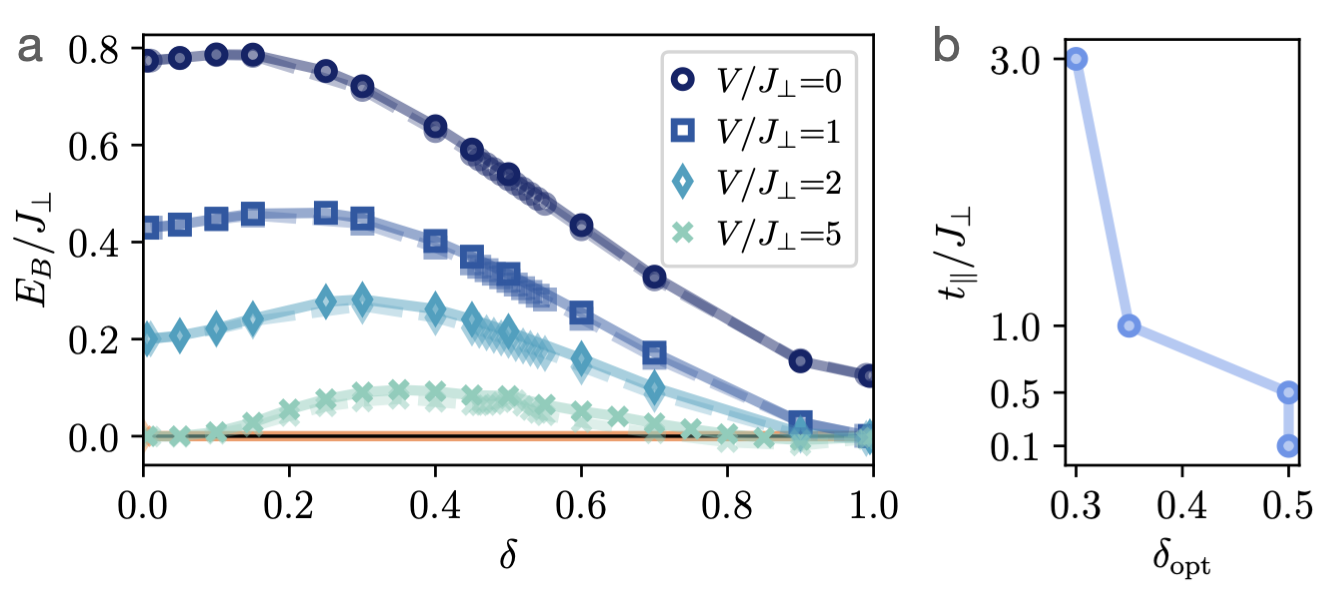}
\caption{a) Binding energies $E_B$ in units of $J_\perp$ for equal hole doping in each chain and with one extra hole added to each chain (blue). Starting from the same state with equal numbers of holes in each chain but adding two extra holes in the same chain, we obtain zero binding energy (orange line). We used a system of length $L_x=100$ (light colors, dashed lines) and $L_x=200$ (solid lines), assumed $t_\parallel/J_\perp=1.0$ and increasing repulsion $V$ for brighter colors. The maximum of the doping dome is found at $\delta_\mathrm{opt}\approx0.3$ for large $V/J_\perp=5$. In b) we show the location of maxima of the doping dome $\delta_\mathrm{opt}$ for $V/J_\perp=5$ but different hopping strengths $t_\parallel/J_\perp$.} 
\label{fig:4}    
\end{figure}

In Fig.~\ref{fig:4} we analyze the binding energies
\begin{align}
        E_B(N_h)=&2\left(E_{N_h-1}-E_{N_h-2}\right) -\left(E_{N_h}-E_{N_h-2}\right)\, ,
\end{align}
with $N_h$ the number of holes doped into the system. We assume that holes are added to the two chains in an alternating fashion, i.e. the state at $N_h+2$ is obtained by adding one hole in each chain. Per definition, positive binding energies indicate that the system tends to form pairs, with stronger binding for larger values of $E_B$. 

In agreement with the chargon-chargon picture of tightly bound hole pairs \cite{Bohrdt2022,Hirthe2023}, we find that $E_B$ for holes in opposite legs (blue lines) is large for $V\leq J_\perp$, and decreases with the hole density of the system $\delta$. In the high doping limit, $E_B$ is small, since there is no magnetic background that stabilizes binding.

%-----------------------------------------
%-------------------------------------------

For $V\leq2J_\perp$, the binding energies show an almost monotonous behaviour with doping. For larger repulsion strengths however, this changes to a non-monotonous, dome-like dependence of $E_B$ on doping, with suppressed $E_B$ in the low and high doping regimes, but enhanced binding for the intermediate doping regime around $\delta_\mathrm{opt}\approx 0.3$ for $t_\parallel/J_\perp=1$. 

A comparison with the density correlations and sc densities from the previous sections shows that the resulting dome of $E_B$ arises in the regime of $V$ where the system forms sc-like pairs rather than cc's. Furthermore, the effective attraction in the finite doping regime and at large $V$ is in agreement with the phenomenology of the Feshbach-mediated pairing mechanism that is expected from the effective sc model Eq. \eqref{eq:H_eff_sc}, describing point-like sc's that experience an effective nearest-neighbor attraction mediated by virtual recombination processes into the cc channel: Near the Feshbach resonance, the effective sc model Eq. \eqref{eq:H_eff_sc} suggests dominant attractive interactions with a maximum of the binding energies in the intermediate doping regime $\delta_\mathrm{opt}\approx0.5$, where the number of neighboring sc's (and hence also the contribution by the nearest-neighbor attraction in Eq. \eqref{eq:H_eff_sc}) is maximal, in agreement with numerical simulations of the effective sc model \eqref{eq:H_eff_sc}, see Fig. \ref{fig:E_sc} in the Appendix. 

For larger hopping strengths $t_\parallel/J_\perp$, binding is stabilized and the resonance shifts to larger $V_c>J_\perp$, yielding effective attraction in the finite doping regime even at $V>J_\perp$, see Fig. \ref{fig:E_sc} in the Appendix. Furthermore, spinon-chargon pairs develop an internal spatial structure, see Fig. \ref{fig:hole_distance}, shifting the maximum to smaller hole dopings, as confirmed by the numerics of the full mixD model Eq. \eqref{eq:mixD_Hamiltonian} shown in Fig. \ref{fig:4}b, e.g. to $\delta_\mathrm{opt}\approx 0.3$ in Fig. \ref{fig:4}a for $t_\parallel/J_\perp=1.0$. In this case, hybridization of the spatially extended sc pairs with cc pairs can lead to an earlier proliferation of cc states, helping mediate strong attractive interactions.

We would like to point out that strong binding in the limit of large doping is consistent with experiments on bilayer nickelates, where superconductivity has been observed at quarter filling, $\delta = 0.5$ \cite{Sun2023}. We find from our analysis that although the binding energies become smaller for larger repulsion strength, binding between holes on opposite legs of the ladder is surprisingly robust; e.g. $E_B\approx 10\% J_\perp$ for a repulsion that is five times larger than all energy scales of the system. In contrast, the binding energy for holes added in the same chain of the ladder (see Fig. \ref{fig:4}a, orange lines) are extremely close to (numerically consistent with) zero for all values of $V$ and $\delta$.

\subsection{Central charge, spin and charge gaps}
\begin{figure}[t]
\centering
\includegraphics[width=0.5\textwidth]{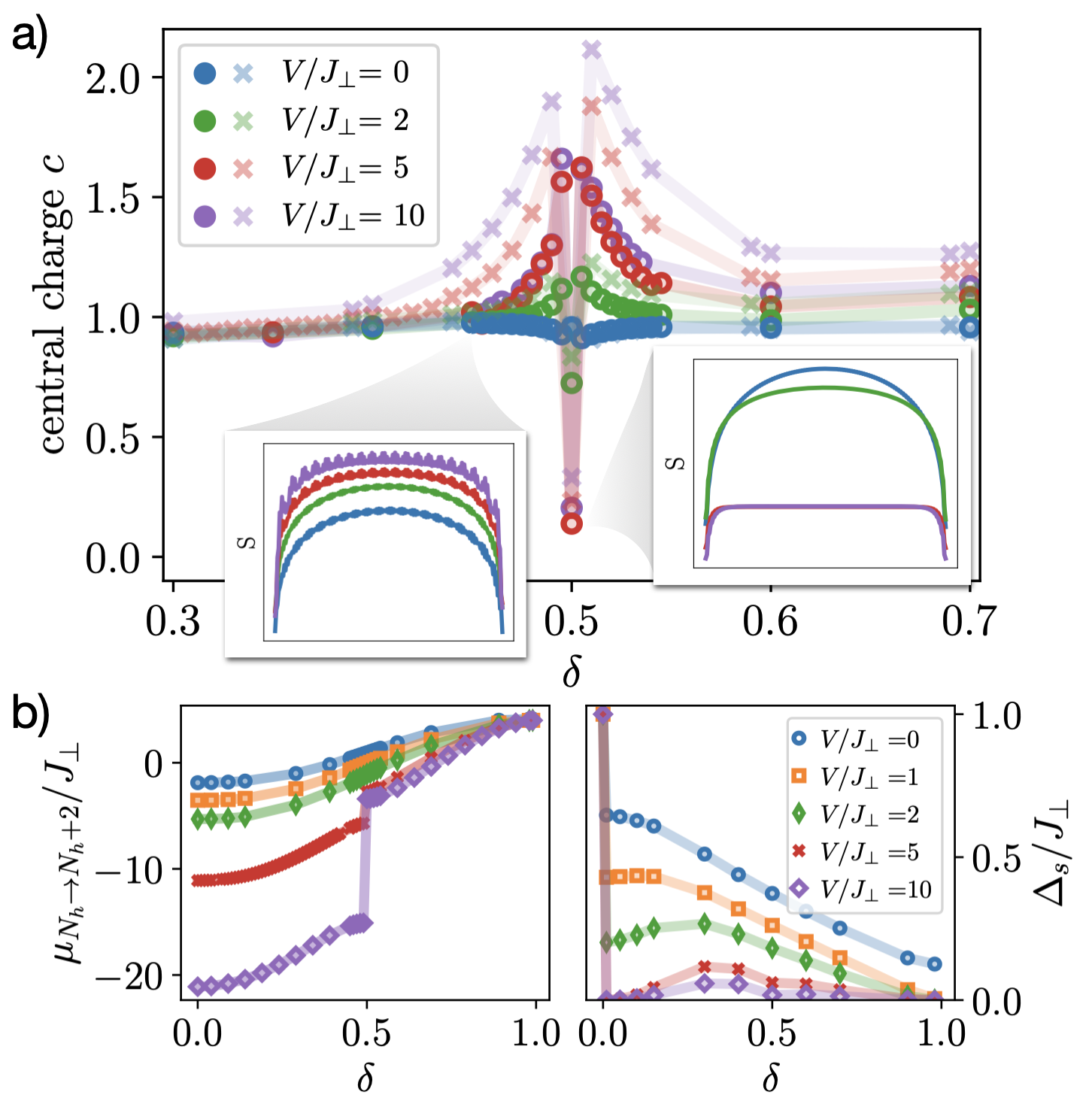}
\caption{a) Central charge extracted from the bipartite entanglement entropy (shown exemplary for $\delta=0.4,\, 0.5$ in the inset figures) by fitting the CFT prediction \eqref{eq:CFT} for $t_{\parallel}/J_\perp=1.0$ and $L_x=100$ (light colors) and $L_x=200$. b) Left: $\mu_{N_h\to N_h+2}=E(N_h+2)-E(N_h)$. At $\delta=0.5$ a charge gap opens for large $V$. Right: Spin gap $\Delta_s = E(S_z^\mathrm{tot}=1)-E(S_z^\mathrm{tot}=0)$. Both calculations were performed for a system of $L_xL_y=200$ sites and $t_{\parallel}/J_\perp=1.0$.} 
\label{fig:5}    
\end{figure}

In the large $V$ limit, two holes on the same rung are strongly penalized and hence no more than one sc can occupy each rung. At commensurate filling $\delta=50\%$, this leads to a charge gap, corresponding to the energy required to create a cc,  which is visible by a jump in 
\begin{align}
    \mu_{N_h\to N_h+2}=E(N_h+2)-E(N_h)
\end{align}
for the respective repulsion strengths at $\delta=0.5$ and $V\geq 5 J_\perp$, see Fig. \ref{fig:5}b. Away from $\delta=0.5$, the charge gap vanishes and the system forms a Luther-Emery liquid without charge gap, but with a spin gap
\begin{align}
    \Delta_s = E(S_z^\mathrm{tot}=1)-E(S_z^\mathrm{tot}=0)
\end{align}
($E$: ground state energy) that remains from the singlet occupation of rungs at $V=0$ \cite{Bohrdt2022} and decreases with increasing $V$.

The Luther-Emery state away from commensurate filling $\delta\neq0.5$ and the opening of the charge gap at $\delta=0.5$ are also reflected in the central charge that can be calculated from the bipartite entanglement entropy $S(x)$, indicating the number of gapless excitations in the system. Numerically, $S$ can be obtained from matrix product states by cutting the system into two parts at bond $x$, since, for an appropriate choice of the MPS chain, the bipartite entanglement entropy between the two parts of the system is entirely carried by a single MPS bond. As can be seen in Fig. \ref{fig:6} and will be discussed later, the density shows significant oscillations, for which we account for by normalizing $S(x)$ by \cite{Palm2022}
\begin{align}
    \Tilde{S}(x)=\frac{2S(x)}{n(x-\frac{1}{2})+n(x+\frac{1}{2})}\bar{n}
\end{align}
with 
\begin{align}
    n(x) = \frac{1}{L_y}\sum_{\mu=1}^{L_y}\langle \hat{n}_{x,\mu}\rangle \quad \mathrm{and}\quad \bar{n}=\frac{1}{L_x}\sum_{x=1}^{L_x}n(x)\,. 
\end{align}
Exemplary results are shown in Fig. \ref{fig:5}a in the insets for $\delta=0.4$ and $\delta=0.5$. 

In order to determine the central charge $c$, we fit the bipartite entanglement entropy to the CFT prediction \cite{Palm2022, Calabrese2004}
\begin{align}
    S_\mathrm{CFT}(x)=\frac{c}{6}\mathrm{log}\left[\frac{2L_x}{\pi}\mathrm{sin}\left(\frac{\pi x}{L_x} \right)\right]+g\,,
    \label{eq:CFT}
\end{align}
where $c$ and $g$ are determined by fitting $S$ to the numerical results. As can be seen in the inset figures of Fig. \ref{fig:5}a, the entanglement entropy flattens for the whole system except for the boundaries when increasing $V$ to a critical value larger than $2J_\perp$ for $\delta=0.5$. In contrast, at $\delta\neq 0.5$ the change is not clearly visible from bare eye.

%-----------------------------------------
%-------------------------------------------

The central charge $c$ obtained from the fits is shown in Fig. \ref{fig:5}a for $t_\parallel/J_\perp=1$ and two system sizes $L_x=100,200$. In a wide range of hole dopings $\delta$, $c=1$, which is in agreement with the fact that there is a finite spin gap. Right at $\delta=0.5$, where the BODW forms for large repulsion strength, the central charge drops to a value close to zero for large enough $V/J_\perp\gtrapprox5$ and both considered system sizes, coinciding with the emergence of the charge gap. Both observations indicate the lack of long-range single spin or single charge order of the emergent phase. We will show later on in Sec. \ref{sec:Friedel} and Sec. \ref{sec:JJSS} that the BODW instead exhibits a plaquette order with plaquettes consisting of two correlated sc's, i.e. two spin and two charge excitations, that form singlets in the spin and charge sector. 

Around $\delta=0.5$, the extracted central charge increases. This is a finite size effect, as can be seen from the drastic change when comparing the results for $200$ and $400$ sites. The strong impact of the size on $c$ can be understood by the fact that the BODW emerges at $\delta=0.5$ and, as shown in Sec. \ref{sec:Friedel}, doping away from $\delta=0.5$ yields domain wall excitations from the bond order. However, the number of excitations $N_{dw}$ is very small compared to the system size, e.g. for a system with $200$ sites and $\delta=0.45$ the number of domain walls is only $N_{dw}=20$. Consequently, extremely large system sizes are needed to decrease this finite size effect.

Our observations on the central charge as well as the charge and spin gaps show that the system forms a Luther-Emery liquid for $\delta \neq 0.5$. As discussed in previous works on atomic BEC to BCS crossovers (e.g. \cite{Recati2005,Tokatly2004,Citro2005}), a Luther-Emery to BEC crossover is continuous and hence difficult to observe, and is further complicated by the quasi-1D setting that we investigate, making it challenging to directly observe the crossover. Nevertheless, we find that the open and closed channel dominated regimes, with cc and $({\rm sc})^2$ constituents respectively, have distincly different characteristics.

\subsection{Isospin and Spin Oscillations \label{sec:JJSS}}
\begin{figure}[t]
\centering
\includegraphics[width=0.5\textwidth]{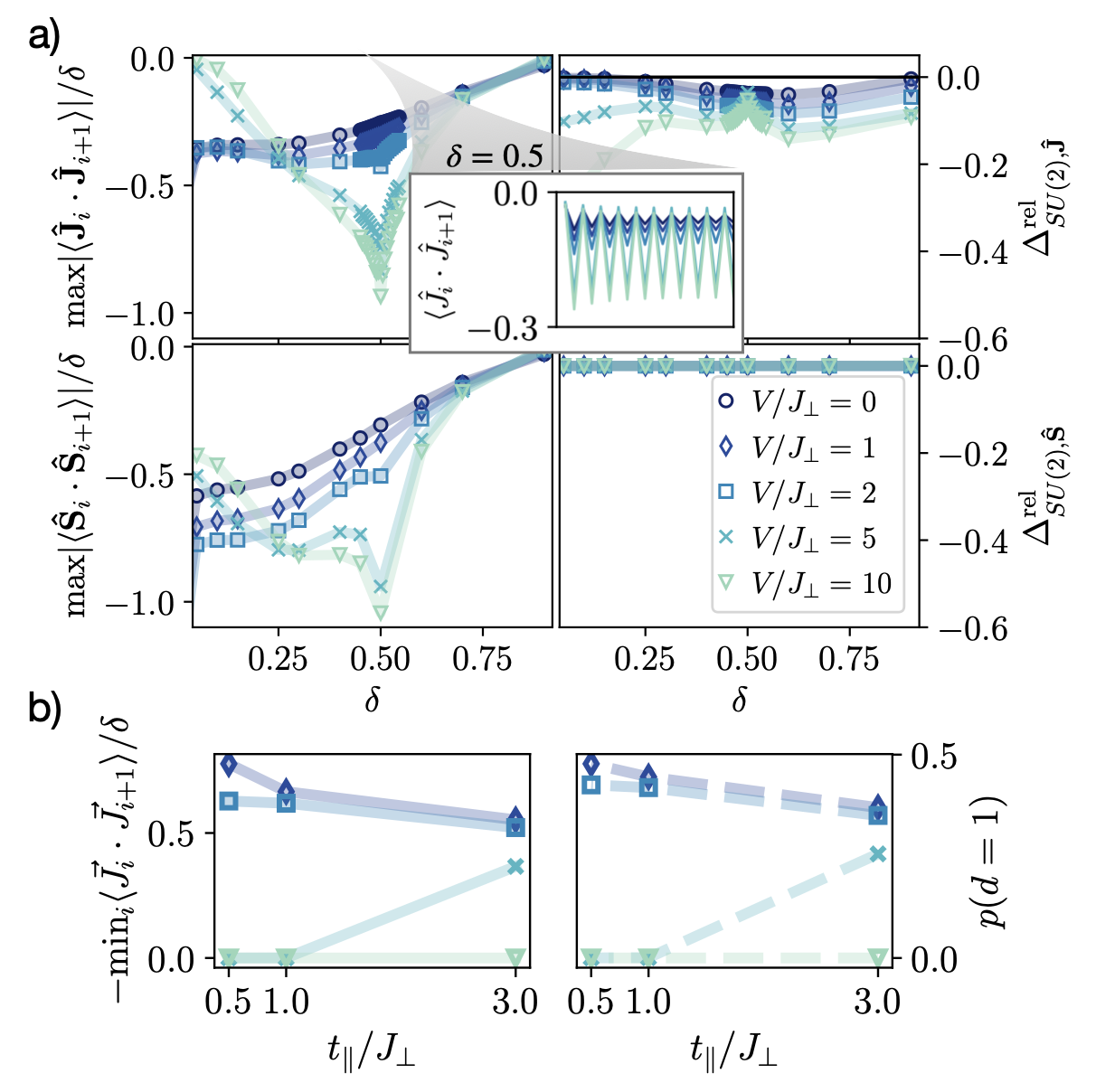}
\caption{a) Minimal value of $\langle \vec{\hat{J}}_i\cdot \vec{\hat{J}}_{i+1}\rangle$ (top left) and \mbox{$\langle \vec{\hat{S}}_i\cdot \vec{\hat{S}}_{i+1}\rangle$} (bottom, left) for $t_\parallel/J_\perp=1.0$. The oscillations of the respective quantities for the first $20$ sites are shown exemplary for $\langle \vec{\hat{J}}_i\cdot \vec{\hat{J}}_{i+1}\rangle$ at $\delta=0.5$ in the inset. The right panel shows $\Delta^\mathrm{rel}_{SU(2),\vec{\hat{J}}/\vec{\hat{S}}}$ as defined in Eq. \eqref{eq:Delta}, that is used as a measure for the singlet character of the plaquettes. b) Minimal value of $\langle \vec{\hat{J}}_i\cdot \vec{\hat{J}}_{i+1}\rangle$ (left) compared to the probability of finding two holes on neighboring rungs $p(d_x=1,d_y=1)$ (right), for $N_h=2$ and different values of $t_\parallel$. All calculations are performed for a system of $400$ sites.} 
\label{fig:8}    
\end{figure}

The special singlet-no singlet order of the BODW is reflected in oscillations of $\langle \vec{\hat{J}}_i\cdot \vec{\hat{J}}_{i+1}\rangle$ as well as $\langle \vec{\hat{S}}_i\cdot \vec{\hat{S}}_{i+1}\rangle$ \cite{lange2023} that are shown in the inset of Fig. \ref{fig:8}a, with minima and maxima arising from the singlet-no singlet alternation, respectively. The left panels of Fig. \ref{fig:8}a show that the absolute value of the minima is the largest for large $V$ and $\delta=0.5$, exactly where the BODW is expected. However, it can be seen that the correlated sc state survives also away from $\delta=0.5$.

Furthermore, we consider the difference between the $x$, $y$ and $z$ components of the isospin $\vec{\hat{J}}_i\cdot \vec{\hat{J}}_{i+1} $, 
\begin{align}
    \Delta^\mathrm{rel}_{SU(2),\vec{\hat{J}}}= \frac{\mathrm{max}\vert \langle \hat{J}^z_i \hat{J}^z_{i+1}\rangle\vert -\mathrm{max}\vert \frac{\langle \hat{J}^+_i \hat{J}^-_{i+1}+\hat{J}^-_i \hat{J}^+_{i+1}\rangle}{4}\vert} {\mathrm{max}\vert \langle \vec{\hat{J}}_i\cdot \vec{\hat{J}}_{i+1}\rangle\vert },
    \label{eq:Delta}
\end{align}
as a measure for the $SU(2)$ symmetry of the $\hat{\vec{J}}$ isospin sector. The results are presented in  Fig. \ref{fig:8}a, top panel and for the spin sector in the bottom panel, with $\Delta^\mathrm{rel}_{SU(2),\vec{\hat{S}}}$ defined analogously. Since the mixD+V model \eqref{eq:mixD_Hamiltonian} is $SU(2)$ symmetric we find $\Delta^\mathrm{rel}_{SU(2),\vec{\hat{S}}}=0$ for all values of $\delta$. This is not the case for the isospin and consequently $\Delta^\mathrm{rel}_{SU(2),\vec{\hat{J}}}\neq 0$. However, we see that for large repulsion strengths $\Delta^\mathrm{rel}_{SU(2),\vec{\hat{J}}}$ is strongly suppressed at $\delta=0.5$ where sc's form isospin singlets on every second site, corresponding to the maxima of $\vert \langle \vec{\hat{J}}_i\cdot \vec{\hat{J}}_{i+1}\rangle \vert$.

%-----------------------------------------
%-------------------------------------------

For low doping and small repulsion, $\langle \vec{\hat{J}}_i\cdot \vec{\hat{J}}_{i+1}\rangle$ and $\langle \vec{\hat{S}}_i\cdot \vec{\hat{S}}_{i+1}\rangle$ are both found to be non-zero as well, but with a smaller value than in the BODW. This, however, is an effect of the small size of the hole pairs in this repulsion regime. If two holes sit on the same rung, there is a finite probability to tunnel to the next site, resulting in a configuration that resembles the BODW. The probability $p(d=1)$ to find pairs of this type on neighboring sites depends on the hopping strength $t_\parallel$ and is shown in Fig. \ref{fig:8}b on the right for the very low doping case of $2$ holes. It can be seen that for small $V$ the probability to find these configurations is indeed higher than for large $V$. The same tendency is reflected in $\langle \vec{\hat{J}}_i\cdot \vec{\hat{J}}_{i+1}\rangle$ and $\langle \vec{\hat{S}}_i\cdot \vec{\hat{S}}_{i+1}\rangle$, see Fig. \ref{fig:8}b left, i.e. the oscillations of $\langle \vec{\hat{J}}_i\cdot \vec{\hat{J}}_{i+1}\rangle$ and $\langle \vec{\hat{S}}_i\cdot \vec{\hat{S}}_{i+1}\rangle$ can be attributed to the relatively large hopping strengths $t_\parallel\approx J_\perp$ used in the DMRG simulations, in contrast to $t_\parallel \ll J_\perp$ assumed in the derivation of Eq. \eqref{eq:H_eff_sc}. For large values of $V$ the holes become largely separated and hence $p(d=1)$ as well as $\langle \vec{\hat{J}}_i\cdot \vec{\hat{J}}_{i+1}\rangle$ and $\langle \vec{\hat{S}}_i\cdot \vec{\hat{S}}_{i+1}\rangle$ are vanishingly small in the low doping limit.

\subsection{Friedel oscillations \label{sec:Friedel}}
\begin{figure}[t]
\centering
\includegraphics[width=0.5\textwidth]{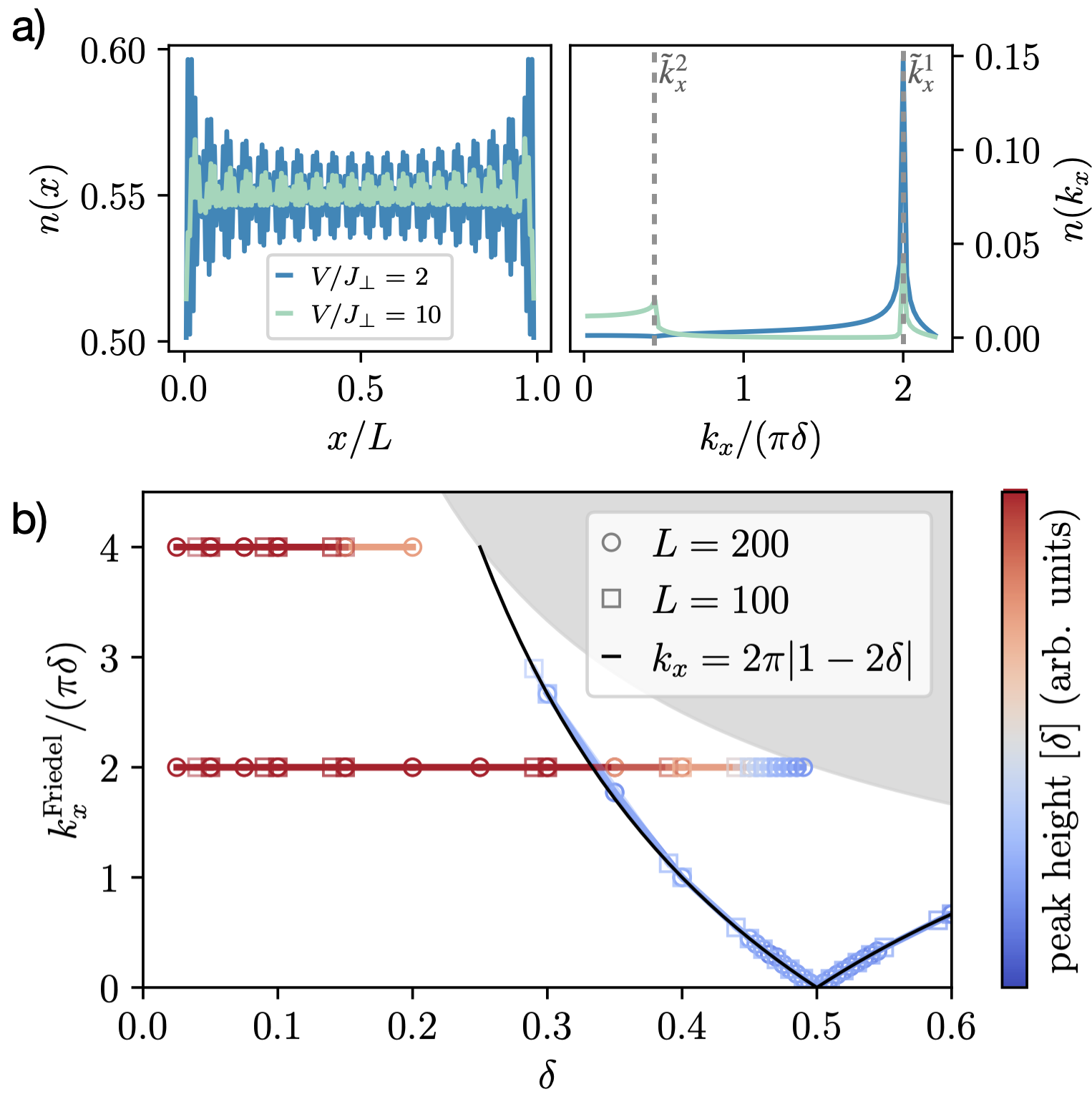}
\caption{a) Friedel oscillations of the density $n(x)$ (left) and its Fourier transforms (right) for a system of length $L_x=200$, $\delta=0.45$ ($N_h=180$) and $t_{\parallel}/J_\perp=1.0$, revealing two peaks at $\tilde{k}_x^1=2\pi\delta$ and $\tilde{k}_x^2=\tilde{k}_x^2(\delta)$ for large $V/J_\perp$. b) The location of the peaks with doping $\delta$, for $L_x=100$ (squares) and $L_x=200$ (circles). The black line is the Friedel peak location of domain walls of the BODW, given by $\tilde{k}_x=2\pi n_{dw}$ with the domain wall density $n_{dw}$ from Eq. \eqref{eq:n_dw}.} 
\label{fig:6}    
\end{figure}

The BODW of correlated sc's allows to understand further observations in the finite doping regime. Here, we focus on the wave vectors $k_x^{\mathrm{Friedel}}$ of Friedel oscillations, extracted from the Fourier transformed density for $L_x=200$ and $t_\parallel/J_\perp=1$. Fig.~\ref{fig:6}a shows the Friedel oscillations of the density (left) and its Fourier transforms (right) for two exemplary values of $V$ and $\delta=0.45$. For small repulsion strength, the Friedel spectrum is dominated by a single peak at $\tilde{k}_x^1=2\pi\delta=2k_F$, where $k_F$ is the Fermi momentum. In view of the finite binding energies, the $\tilde{k}_x^1$-peak can be understood as arising from a liquid composed of either individual cc's or pairs of sc's ($(\mathrm{sc})^2$), both corresponding to a density $n_{cc}=\frac{1}{2} N_h/L=\delta$ or $n_{sc^2}=N_h/(2 L)=\delta$. The doping dependence of $\tilde{k}_x^1$ is shown in Fig. \ref{fig:6}b, where it can be seen that this peak, independent of doping, survives up to the resolution limit from the finite lattice spacing of the system denoted by the gray area. Furthermore, Fig. \ref{fig:6}b shows that there is a higher harmonic of $\tilde{k}_x^1$, possibly arising due to residual short-range repulsion. 

%-----------------------------------------
%-------------------------------------------

In addition, a second, smaller peak arises only for large $V$ with $k_x=0.44\pi\delta$ for the considered system with hole doping $\delta=0.45$. As shown in Fig. \ref{fig:6}b, this second peak has a markedly different doping dependence than $\tilde{k}_x^1$, with decreasing $\tilde{k}_x^2$ up to $\delta=0.5$ and increasing $\tilde{k}_x^2$ for $\delta>0.5$. We argue that this peak is associated with excitations of the BODW found at $\delta=0.5$. The latter correspond to domain walls of the plaquette order, with rung-singlet ($\delta<0.5$) or cc ($\delta>0.5$) character, whose density is
\begin{align}    n_{dw}=\vert 1-2\delta\vert .
\label{eq:n_dw}
\end{align}
Respective peaks arise at $\tilde{k}_x=2\pi n_{dw}$ (black line), which is in excellent agreement with the location of the peak in the whole density regime that was calculated numerically for $L_x=100$ and $L_x=200$, see \ref{fig:6}b. This supports our observations from before that in the intermediate doping regime, the system can be understood in terms of correlated spinon-chargon pairs, forming a BODW state of Cooper pairs with $(\mathrm{sc})^2$ character and domain walls between them.

\section{Summary and Outlook}
To conclude, we have presented the mixD+V model with repulsive Coulomb interactions $V$ -- closely connected to the newly discovered bilayer nickelate superconductor La$_3$Ni$_2$O$_7$ \cite{Sun2023,Qu2023} -- as a setting to investigate binding in strongly correlated quantum systems from the perspective of BEC to BCS crossovers and Feshbach-mediated pairing. Our numerical and analytical results show that binding in open channel dominated state, on the BCS side of the Feshbach resonance, is mediated by coupling to the the closed chargon-chargon channel familiar from Feshbach resonances. This opens the way for a more detailed analysis of Feshbach-mediated pairing, in the context of mixD ladders, bilayers, nickelates and beyond \cite{Crepel2023}. 

The mixD setup allows a controlled crossover from a clearly BEC-like pairing mechanism with tightly bound holes for $V=0$ to a state of correlated spinon-chargon pairs at strong repulsion $V$ resembling closer a BCS state of fermions. We see very clear signals for a drastic change of the nature of the charge carriers in the numerical results of the low and high repulsion limits, e.g. in correlation functions and binding energies. In contrast, the central charge and Friedel oscillations remain unchanged, as expected from a Luther-Emery liquid. Moreover, our comparison to effective descriptions in terms of tightly bound chargon-chargon pairs for low repulsion and spatially extended bound states of spinon-chargon pairs for high $V$ indicate that the system can indeed be described by very different types of charge carriers in the respective regimes. In order to investigate this change further, spectroscopic probes could be used to distinguish the excitations in both regimes by their dispersion. Lastly, the mixD+V ladder can also be investigated using ultracold atom experiments, where the nearest-neighbor repulsion can be realized by doping the setup of Ref. \cite{Hirthe2023} with holes and doublons on opposite legs of the ladder. Alternatively, our model can be probed in tweezer arrays of Rydberg atoms or ultracold polar molecules realizing a tunable \mbox{$t$-$J$-$V$} Hamiltonian with hard-core bosonic holes~\cite{Homeier2023_bosonictJ} since the physics remains unaffected by the particle statistics in the mixD+V ladder geometry. Altogether, this may contribute to the search for materials with high superconducting transition temperatures~\cite{Hur2009,Hur2011}.

%%%%%%%%%%%%%%%%%%%%%%%%%%%%%

% % % % % % % % % % % % % % % % % % % % % % % % % % % 

\emph{Note added.--}While finishing the manuscript, we became aware of a closely related work by H.~Yang, H.~Oh and Y.~H.~Zhang~\cite{Yang2023-arXiv}, in which they use the DMRG method to study a similar bilayer repulsive~$t$-$J$ model on a two-leg ladder. In their work, they also find the emergence of Feshbach resonance and propose a doping induced BEC-to-BCS crossover scenario for the bilayer nickelates.

%%%%%%%%%%%%%%%%%%%%%%%%%%%%%%%%%%%%%%
\emph{Acknowledgements.--} We would like to thank Atac Imamoglu, Daniel Jirovec, Felix Palm, Henning Schlömer, Immanuel Bloch, Ivan Morera Navarro, Lieven Vandersypen, Markus Greiner, Matjaz Kebric, Pablo Cova Farina, Tim Harris and Tizian Blatz for helpful discussions. Special thanks to Henning Schl\"omer for his help with the DMRG implementation of the mixD symmetries. We acknowledge funding by the Deutsche Forschungsgemeinschaft (DFG, German Research Foundation) under Germany's Excellence Strategy -- EXC-2111 -- 390814868 and from the European Research Council (ERC) under the European Union’s Horizon 2020 research and innovation programm (Grant Agreement no 948141) — ERC Starting Grant SimUcQuam. ED acknowledges support from the ARO grant W911NF-20-1-0163 and the SNSF project 200021-212899. HL acknowledges support by the International Max Planck Research School. LH acknowledges support by Studienstiftung des deutschen Volkes.

%%%%%%%%%%%%%%%%%%%%%%%%%%%%%%%%%%%%%%%%
%merlin.mbs apsrev4-1.bst 2010-07-25 4.21a (PWD, AO, DPC) hacked
%Control: key (0)
%Control: author (72) initials jnrlst
%Control: editor formatted (1) identically to author
%Control: production of article title (-1) disabled
%Control: page (0) single
%Control: year (1) truncated
%Control: production of eprint (0) enabled
%

\clearpage
\newpage~
%~\newpage

\appendix
\onecolumngrid

\section*{Appendix}

\subsection{Brief review: binding in the mixD model without repulsion \label{appendix:mixDbinding}}
\begin{figure}[htp]
\centering
\includegraphics[width=0.8\textwidth]{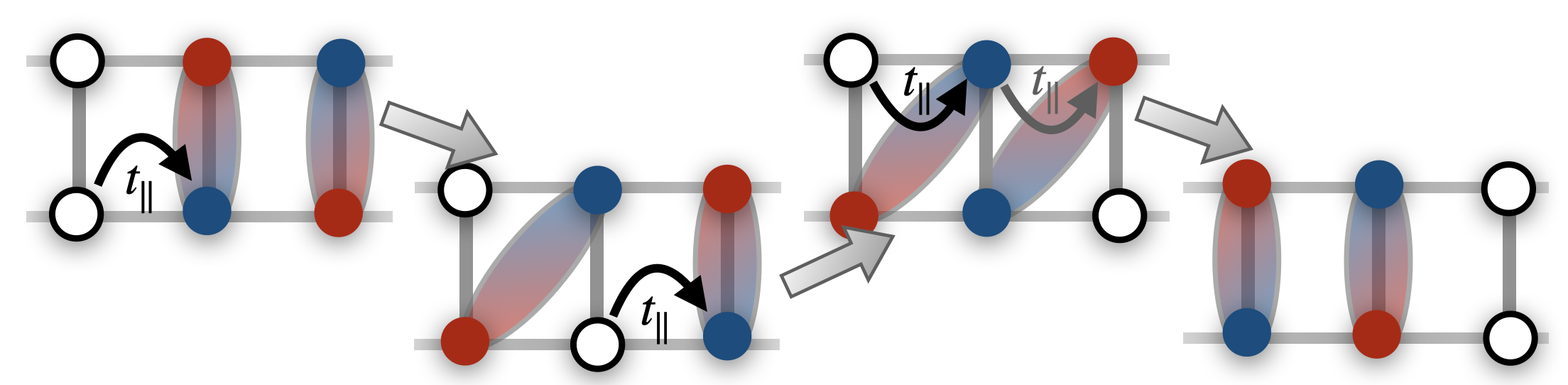}
\caption{Schematic illustration of the binding mechanism in the mixD ladder: a) A single, unbound hole moves through the ladder and distorts the spin background.  b) Retracing mechanism of two holes, yielding a restored singlet background when the holes move through the system together. } 
\label{fig:9}    
\end{figure}

As shown in Refs. \cite{Bohrdt2022,Hirthe2023} the strong binding energies observed in the mixD model result from two peculiarities of the model:
\begin{enumerate}
    \item The ladder: Compared to a single hole that distorts the background singlet order when moving through the ladder (see Fig. \ref{fig:9}a), a second hole can retrace the first one and restore the background order, making it favorable for the system to pair holes. This is schematically shown in Fig. \ref{fig:9}b.
    \item The mixed dimension: Suppressing the hopping term between the chains enhances the probability for two holes to sit on the same rung since the effect of Pauli blocking is suppressed. This is further discussed in Ref. \cite{Hirthe2023}. 
\end{enumerate}

\subsection{Experimental Realization of the mixD ladder with repulsive interactions \label{appendix:Experiment}}

The mixD Hamiltonian \eqref{eq:mixD_Hamiltonian} without repulsive interactions $V$ has already been realized by Hirthe et al. \cite{Hirthe2023}, by applying potential offset $\Delta$ between the upper and lower leg of a Fermi Hubbard ladder with large $U\gg\Tilde{t}_\parallel, \Tilde{t}_\perp$. For $U>\Delta \gg\Tilde{t}_\perp$ tunneling between the chains is suppressed to an effective $t_\perp=0$ and gives rise to a spin-exchange
\begin{align}
    J_\perp = \frac{2\Tilde{t}_\perp^2}{U+\Delta}+\frac{2\Tilde{t}_\perp^2}{U-\Delta}=4\Tilde{t}_\perp^2\frac{U}{U^2-\Delta^2}\geq 0\,.
\end{align}
If the chains are occupied only by single spins and holes (no doublons), the virtual tunneling between site $i$ in the upper layer and $j$ in the lower layer is given by $-\frac{\Tilde{t}_\perp^2}{\Delta}(1-\hat{n}_i)\hat{n}_j$ (for a hole at site $i$ in the upper layer and spin at site $j$ in the lower layer) or $+\frac{\Tilde{t}_\perp^2}{\Delta}\hat{n}_i(1-\hat{n}_j)$ (for a hole at site $i$ and spin at site $j$) respectively. In total, this leads to 
\begin{align}
\hat{\mathcal{H}}_{\mathrm{eff}}^{h}= \sum_{ i }J_\perp\hat{\vec{S}}_{i,1}\cdot \hat{\vec{S}}_{i,0} -\frac{J_\perp}{4}\hat{n}_{i,1}\hat{n}_{i, 0}-\frac{\Tilde{t}_\perp^2}{\Delta}(\hat{n}_i-\hat{n}_j).
\end{align}
Similarly, if we have doublons in both layers one gets contributions by virtual tunnelings between two doublons or a doublon and a single spin at sites $i$ and $j$ in opposite layers (no holes in the system). In this case, we find $+\frac{\Tilde{t}_\perp^2}{\Delta}(1-\hat{\Tilde{n}}_i)\hat{\Tilde{n}}_j$ (for a doublon at site $i$ in the upper layer and spin at site $j$ in the lower layer) or $-\frac{\Tilde{t}_\perp^2}{\Delta}\hat{\Tilde{n}}_i(1-\hat{\Tilde{n}}_j)$ (for a spin at site $i$ in the upper layer and doublon at site $j$ in the lower layer) and hence
\begin{align}
    \hat{\mathcal{H}}_{\mathrm{eff}}^{2d}= \sum_{ i }J_\perp\hat{\vec{S}}_{i,1}\cdot \hat{\vec{S}}_{i,0} -\frac{J_\perp}{4}\hat{n}_{i,1}\hat{n}_{i, 0}+\frac{\Tilde{t}_\perp^2}{\Delta}(\hat{\Tilde{n}}_i-\hat{\Tilde{n}}_j)
\end{align}
if we introduce the number operator $\hat{\Tilde{n}}$ with
\begin{align}
    \Tilde{n}_i = \left\{
    \begin{matrix}
    0 \quad \mathrm{if}\, n_i=2 \, \mathrm{(doublons) ~ or ~}n_i=0  \mathrm{~(holes)}\\
    \quad 1 \quad \mathrm{if}\, n_i=1 \, \mathrm{(single}\,\mathrm{particle)}
    \end{matrix} \right.\,.
\end{align}
Both cases correspond to a constant shift and does not change the physics of the mixD ladder.

In contrast, if we add doublons in only one of the chains, i.e. the lower chain, we get 
\begin{itemize}
    \item a contribution $-\frac{\Tilde{t}_\perp^2}{\Delta}(1-\hat{\Tilde{n}}_i)\hat{n}_j$ (for a doublon at site $i$ in the lower leg and spin at site $j$ in the upper leg)
    \item a contribution $-\frac{\Tilde{t}_\perp^2}{\Delta}\hat{\Tilde{n}}_i(1-\hat{n}_j)$ (for a single spin at site $i$ in the lower leg and a hole at site $j$ in the upper leg)
    \item  or $-2\frac{\Tilde{t}_\perp^2}{\Delta-U}(1-\hat{\Tilde{n}}_i)(1-\hat{n}_j)$ (for a doublon at site $i$ in the lower leg and a hole at site $j$ in the upper leg).
\end{itemize}
In total, we have a contribution

\begin{align}  \hat{\mathcal{H}}_{\mathrm{eff}}^{1d}(j)= 
J_\perp \left( \hat{\vec{S}}_{j\mu}\cdot \hat{\vec{S}}_{j\bar{\mu}} - \frac{1}{4} \hat{n}_{j,\mu} \hat{n}_{j,\overline{\mu}} \right) +\left(2\frac{\tilde{t}^2}{\Delta}-2 \frac{\tilde{t}^2}{\Delta-U}\right)\hat{\Tilde{n}}_{j\mu} \hat{\Tilde{n}}_{j\overline{\mu}} 
    +\left( -\frac{\tilde{t}^2}{\Delta}+2\frac{\tilde{t}^2}{\Delta-U}\right)\sum_{ \mu} \hat{\Tilde{n}}_{i\mu} - 2\frac{\tilde{t}^2}{\Delta-U}  \,
\end{align}
and hence
\begin{align}
\hat{\mathcal{H}}_{\mathrm{eff}}^{1d} = J_\perp\sum_{j } \left( \hat{\vec{S}}_{j\mu}\cdot \hat{\vec{S}}_{j\bar{\mu}} - \frac{1}{4} \hat{n}_{j,\mu} \hat{n}_{j,\overline{\mu}} \right) +V\sum_{ i } \hat{\Tilde{n}}_{i\mu} \hat{\Tilde{n}}_{i\overline{\mu}} 
    + \epsilon_0\sum_{ i \mu} \hat{\Tilde{n}}_{i\mu}+\mathrm{const.}\,
\end{align}
where we can define 
\begin{align}
    V-\frac{J_\perp}{4} :=-\frac{J_\perp}{4}+2\frac{\Tilde{t}_\perp^2}{\Delta}+2\frac{\Tilde{t}_\perp^2}{U - \Delta} =\Tilde{t}_\perp^2\left(\frac{2}{\Delta}+\frac{U+2\Delta}{U^2-\Delta^2}\right)
\end{align}
and 
\begin{align}
\epsilon_0:=-\frac{\tilde{t}^2}{\Delta}+2\frac{\tilde{t}^2}{\Delta-U}.
\end{align}
$V$ is repulsive for doublon dopants in the lower chain and hole dopants in the upper chain ($\Delta \geq 0$ and $V\geq 0$) and attractive for doublons in the upper chain and holes in the lower chain ($\Delta \leq 0$ and $V\leq 0$). Note that we have the restrictions $1<\frac{\vert \Delta \vert}{\Tilde{t}_\perp}$, $1<\frac{\vert U \vert}{\Tilde{t}_\perp}$, $\frac{\vert \Delta \vert}{\Tilde{t}_\perp} < \frac{\vert U \vert}{\Tilde{t}_\perp}$ and $1<\frac{\vert U \pm \Delta \vert}{\Tilde{t}_\perp}$ denoted by the black lines in Fig.~\ref{fig:2} in the main text.

\subsection{Details of our numerical DMRG simulations \label{appendix:DMRG}}
We use the single-site density matrix renormalization group (DMRG) algorithm implemented in the package SyTen \cite{syten1,syten2}. The implementation of the mixD model is based on essentially the same as in Schlömer et al. \cite{Schloemer2022}: We explicitly employ $U(1)_{N_{\mu=1}}\otimes U(1)_{N_{\mu=2}}\otimes U(1)_{S_z^\mathrm{tot}}$ -- associated with charge conservation in each individual leg (since $t_\perp=0$) and total magnetization conservation -- in the DMRG ground state calculations ($N_{\mu=i}$: number of particles in chain $i$). As shown in the Appendix of Ref. \cite{Schloemer2022} this makes the ground state search much more efficient compared to calculations with only global charge conservation $U(1)_{N}\otimes  U(1)_{S_z^\mathrm{tot}}$.

\subsubsection{Exemplary convergence tests}
\begin{figure}[t]
\centering
\includegraphics[width=0.8\textwidth]{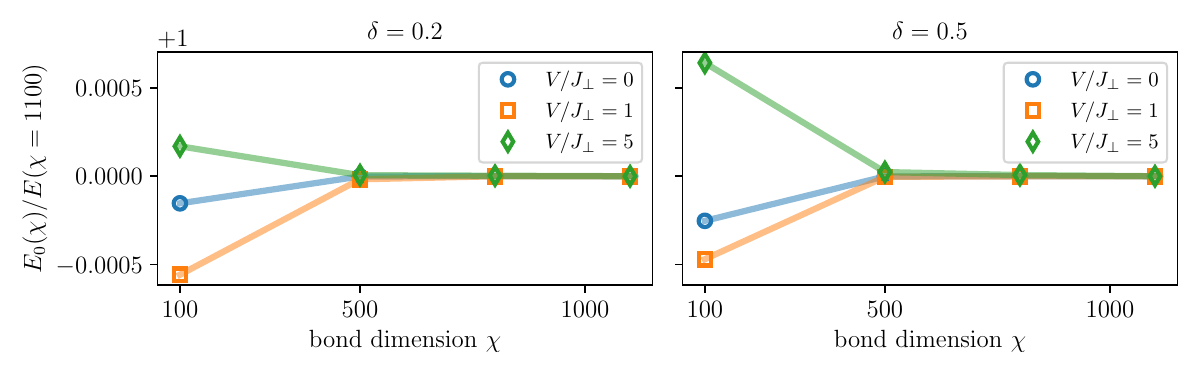}
\includegraphics[width=0.8\textwidth]{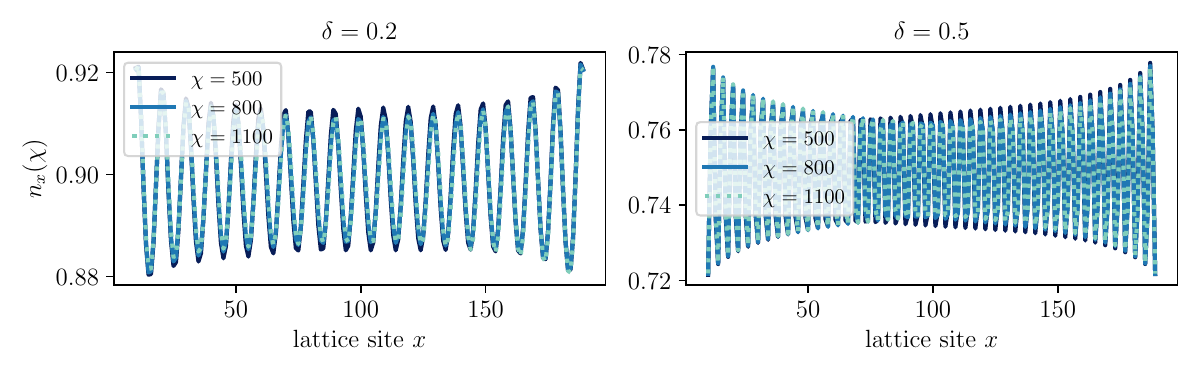}
\caption{We show exemplary convergence tests for a system with $400$ sites and $t_\parallel/J_\perp=1$, as typically applied for the results presented in the main text. Here, we compare ground state energies $E_0(\chi)$ (top) and local densities $n_x(\chi)$ ($V/J_\perp=5$, bottom) for $\chi=100,\dots,1100$. Convergence is typically achieved for $\chi>500$. } 
\label{fig:Convergence}    
\end{figure}

\begin{figure}[t]
\centering
\includegraphics[width=0.8\textwidth]{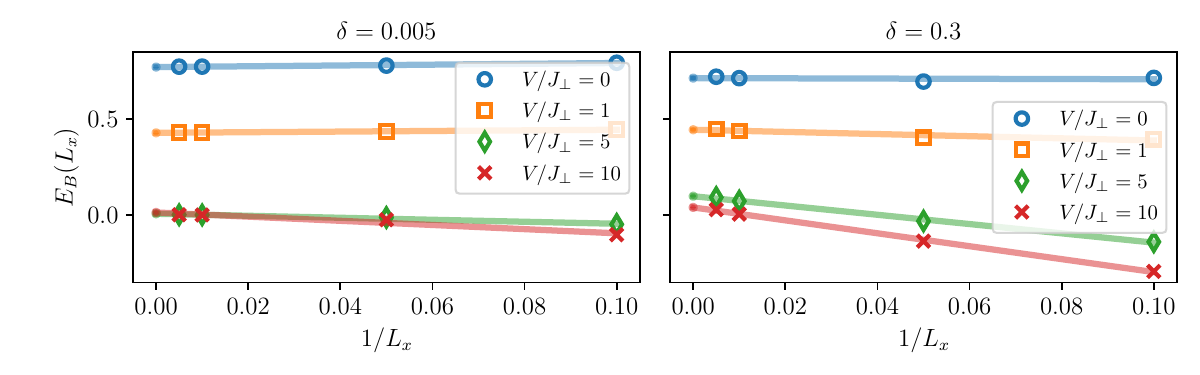}
\caption{Finite size scaling for systems with $L_x=10,\dots,200$ and $t_\parallel/J_\perp=1$, as typically applied for the results presented in the main text.} 
\label{fig:Finite_size}    
\end{figure}

Convergence is ensured by comparing energies, their variance and other expectation values like the density for different bond dimensions $\chi$ up to $\chi_\mathrm{max}=1100$. Typical bond dimensions we use for the results presented in the main text are $\chi \approx 1000$. Examples for a system of length $L_x=200$ with $t_\parallel/J_\perp=1$ and different repulsion strengths $V$ are shown in Fig. \ref{fig:Convergence}. It can be seen that convergence is typically achieved for $\chi>500$, with slightly slower convergence for commensurate filling $\delta=50\%$.

\subsubsection{Binding energies for different hopping strengths}
\begin{figure}[t]
\centering
\includegraphics[width=0.48\textwidth]{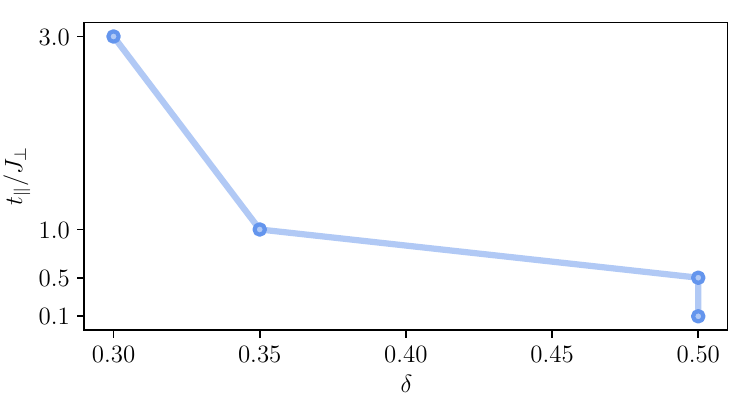}
\includegraphics[width=0.51\textwidth]{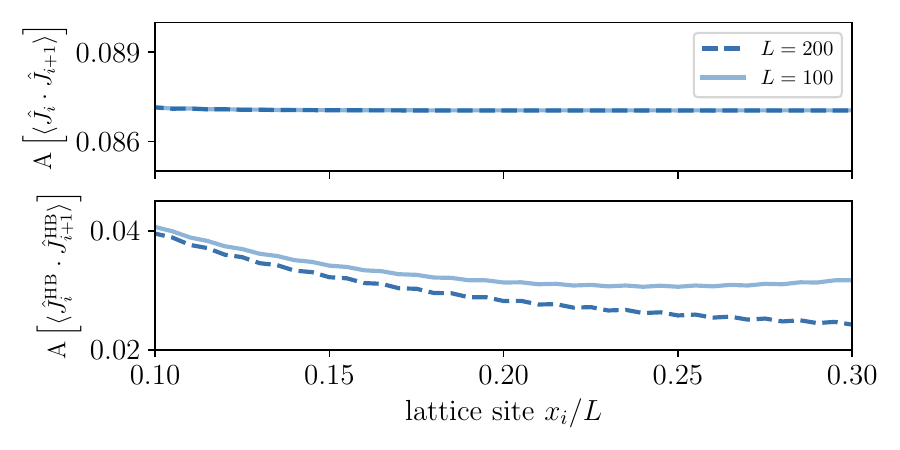}
\caption{Left: We show hole dopings $\delta_\mathrm{opt}$ for which the maximum of the binding energies arises. Different $t_\parallel/J_\perp$ and a system of $L_x=100$ and $V/J_\perp=5$ are considered. Right: Amplitude of the VBS-like oscillations of $\langle \hat{J}_i\cdot \hat{J}_{i+1}\rangle$ for the mixD ladder with $t_\parallel/J_\perp=1$, $V/J_\perp=5$ and length $L_x=100$ and $L_x=200$ (top). We compare the results to a 1D Heisenberg (HB) model of the same form as $\hat{H}_\mathrm{eff}^{J_\perp=0}$ \eqref{eq:Heff_sc_J=0} with $4\frac{t_\parallel^2}{V}=1$ (bottom).} 
\label{fig:EB_max}    
\end{figure}

The maximum of the binding energies presented in Fig. \ref{fig:1}b shifts with the hopping strength. In Fig. \ref{fig:EB_max} on the left, we show the maximum of binding energies $\delta_\mathrm{opt}$ for different hopping strengths, revealing a strong dependence of $\delta_\mathrm{opt}$ on $t_\parallel/J_\perp$. The origin of this shift can be understood by considering the effective sc Hamilronian Eq. \eqref{eq:H_eff_sc} that describes point-like, hard core sc's that interact attractively of they sit on neighboring rungs in the limit of $t_\parallel \gg J_\perp, J_\perp-V$. Consequently, we expect a maximal binding energy at $\delta_\mathrm{opt}=0.5$ in this regime, as confirmed numerically in Fig. \ref{fig:EB_max}. For larger hopping strengths sc's develop a spatial structure and extend over several sites as can be seen in Fig. \ref{fig:hole_distance}. Hence, they interact strongly already for $\delta<0.5$ and $\delta_\mathrm{opt}$ shifts to smaller values.

\subsubsection{$\langle \hat{J}_i\cdot  \hat{J}_{i+1}\rangle$ and $\langle \hat{S}_i\cdot  \hat{S}_{i+1}\rangle$ at $\delta=0.5$}

In the main text we discuss the expectation values $\langle \hat{J}_i\cdot  \hat{J}_{i+1}\rangle$ and $\langle \hat{S}_i\cdot  \hat{S}_{i+1}\rangle$, showing strong oscillations indicating the BODW with alternating singlet - no singlet order at $\delta=0.5$. Here, we provide a more detailed comparison of the Amplitudes of $\langle \hat{J}_i\cdot  \hat{J}_{i+1}\rangle$ shown in Fig. \ref{fig:4}b, see Fig. \ref{fig:EB_max} on the right. For the mixD model, $\langle \hat{J}_i\cdot  \hat{J}_{i+1}\rangle$ and $\langle \hat{S}_i\cdot  \hat{S}_{i+1}\rangle$ show VBS-like oscillations of significant amplitude. This is in agreement with our BODW interpretation, where minima of $\langle \hat{J}_i\cdot  \hat{J}_{i+1}\rangle$ correspond to singlet bonds of the plaquettes, and maxima to no singlets between the plaquettes. To show the robustness of these oscillations we present the results for two different system sizes $L_x=100,\, 200$. Note that the same quantity evaluated for the pure Heisenberg model with $\hat{\vec{J}}$-spins and open boundaries show oscillations of a smaller amplitude than for the mixD case as well. However, these oscillations show a strong dependence on the system size, see Ref. \cite{lange2023} and Fig. \ref{fig:EB_max}. 

\subsubsection{Hole distance in the limit of low doping}
\label{AppdxHoleDistance}
\begin{figure}[t]
\centering
\includegraphics[width=0.48\textwidth]{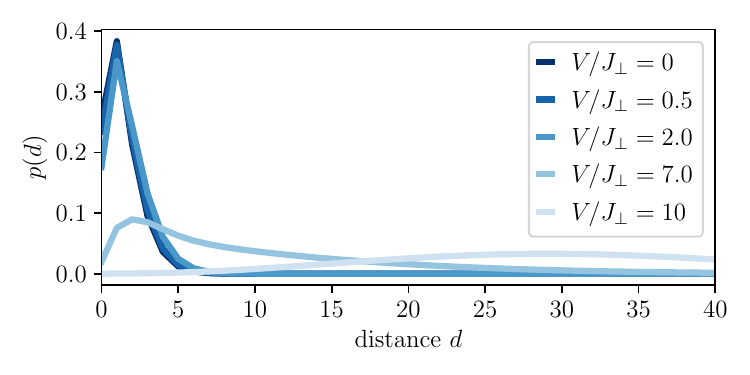}
\caption{We show the hole distance calculated by DMRG in the limit of low doping for $t_\parallel/J_\perp=3$, different repulsion strengths $V$ and $L_x=80$. } 
\label{fig:hole_distance}    
\end{figure}

Furthermore, we mention the average distance of holes in the main text. The average hole distance is calculated using DMRG by evaluating the probability to find the holes / particles at a certain distance $d$, given by \cite{Grusdt2019}
\begin{align}
    p(d)=\sum_{i,j \,\mathrm{s.t.}\,\vert i-j\vert =d} \langle \hat{n}_{i,\mu}\hat{n}_{j,\bar{\mu}}\rangle.
\end{align}
An example is shown in Fig. \ref{fig:hole_distance} for $t_\parallel/J_\perp=3$. One can see that the distribution broadens with increasing $V$. The average hole distance is given by $d_h=\sum_d p(d)$. 

\subsection{Derivation of the effective chargon-chargon model \label{appendix:Derivation_cc_Model}}
In order to derive the effective chargon-chargon Hamiltonian, we introduce the notation
\begin{align}
\hat{b}_{i}^\dagger \ket{\dots \singlet \dots}=\ket{\dots \holes\dots}
\end{align}
and 
\begin{align}
    \hat{b}_{i} \ket{\dots \holes \dots}=\ket{\dots \singlet \dots}\, ,
\end{align}
where singlets are denoted by $\singlet$ and chargon-chargon pairs by $\holes$.

First of all, we notice that we can rewrite $\hat{\mathcal{H}}_J$ in terms of the chargon-chargon operators $\hat{b}_i^{(\dagger)}$, i.e.
\begin{align*}
    \hat{\mathcal{H}}_J &= J_\perp \sum_j \left(\hat{c}_{j,1,\downarrow}^\dagger\hat{c}_{j,1,\downarrow}\hat{c}_{j,2,\uparrow}^\dagger\hat{c}_{j,2,\uparrow}+\hat{c}_{j,1,\uparrow}^\dagger\hat{c}_{j,1,\uparrow}\hat{c}_{j,2,\downarrow}^\dagger\hat{c}_{j,2,\downarrow}-\hat{c}_{j,1,\uparrow}^\dagger\hat{c}_{j,1,\downarrow}\hat{c}_{j,2,\downarrow}^\dagger\hat{c}_{j,2,\uparrow}-\hat{c}_{j,1,\downarrow}^\dagger\hat{c}_{j,1,\uparrow}\hat{c}_{j,2,\uparrow}^\dagger\hat{c}_{j,2,\downarrow}\right)
    \notag \\
    &=J_\perp \sum_j\hat{b}_j\hat{b}_j^\dagger =J_\perp \sum_j(1+\hat{b}_j^\dagger \hat{b}_j)
\end{align*}

The hopping term can be considered by performing a Schrieffer-Wolff transformation and restricting to the energy subspace related to energy scales $V-J_\perp$ of chargon-chargon excitations. We assume that $V\ll J_\perp$, i.e. the chargon-chargon states are the low energy states of the system and their energy subspace is well separated from the energy subspace of spinon-chargon pairs. The transformed Hamiltonian consists includes the correction $\Delta \hat{\mathcal{H}}_t$ given by second order processes via the higher energy spinon-chargon channel, i.e.
\begin{align}
\hat{\mathcal{H}}^{cc}_{eff} &= -2\frac{t_\parallel^2}{J_\perp-V}\sum_{\langle ij\rangle }\hat{\mathcal{P}}_b \left(\hat{b}_i^\dagger \hat{b}_j +\mathrm{h.c.}\right)\hat{\mathcal{P}}_b
    +4\frac{t_\parallel^2}{J_\perp-V}\sum_{\langle ij\rangle } \hat{b}_i^\dagger\hat{b}_i \hat{b}_j^\dagger \hat{b}_j -2\frac{t_\parallel^2}{J_\perp-V}\sum_{\langle ij\rangle } \hat{b}_j^\dagger \hat{b}_j+(V-J_\perp) \sum_j\hat{b}_j^\dagger \hat{b}_j\notag\\
    &=-\underbrace{2\frac{t_\parallel^2}{J_\perp-V}}_{=:t_{eff}^{cc}}\sum_{\langle ij\rangle }\hat{\mathcal{P}}_b \left(\hat{b}_i^\dagger \hat{b}_j +\mathrm{h.c.}\right)\hat{\mathcal{P}}_b
    +\underbrace{4\frac{t_\parallel^2}{J_\perp-V}}_{=:V_{eff}^{cc}}\sum_{\langle ij\rangle } \hat{b}_i^\dagger\hat{b}_i \hat{b}_j^\dagger \hat{b}_j
    -(\underbrace{z2\frac{t_\parallel^2}{J_\perp-V}}_{=:\mu_{eff}^{cc}} +J_\perp -V)\sum_{j }\hat{b}_j^\dagger \hat{b}_j.
    \label{eq:cc_effective_Ham}
\end{align}
Here, $\hat{\mathcal{P}}_b$ is the Gutzwiller projector on states with maximally one boson per site. Note that the second term results from the fact that the energy reduction by the kinetic contribution is smaller (i.e. the energy is higher) if there are two chargon-chargon pairs next to each other. 

If we neglect the last term which gives a constant contribution when assuming a fixed chargon-chargon number we can rewrite $\hat{\mathcal{H}}^{cc}_{eff}$ in terms of spin operators by mapping $Z_i=\hat{b}_i^\dagger \hat{b}_i-\frac{1}{2}$ and $\hat{b}_i^{(\dagger)}$ to the respective spin raising and lowering operators and obtain \cite{Bohrdt2021}
\begin{align}
    \hat{\mathcal{H}}^{cc}_{eff}= -\underbrace{4\frac{t_\parallel^2}{J_\perp-V}}_{=:J_{eff}^{cc}}\sum_{\langle ij\rangle } \left( X_jX_i+Y_jY_i-Z_jZ_i\right).
    \label{eq:XXZ}
\end{align}
This is an XXZ model with $J_{eff}^{cc,xy}=-J_{eff}^{cc,z}<0$ (for $V<J_\perp$).

\subsection{Derivation of the effective spinon-chargon model \label{appendix:Derivation_sc_Model}}

\begin{figure}[t]
\centering
\includegraphics[width=0.7\textwidth]{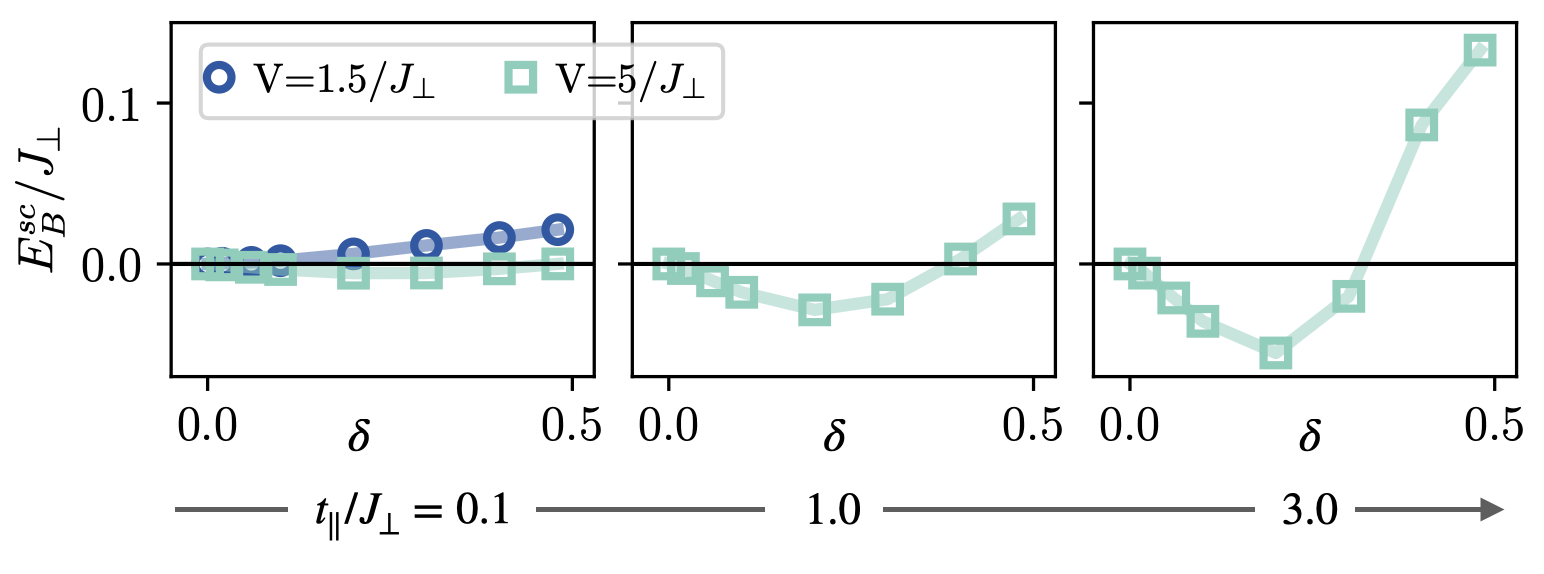}
\caption{Binding energies from DMRG simulations of the effective sc model, Eq. \eqref{eq:H_eff_sc}, for $t_\parallel/J_\perp=0.1, 1.0, 3.0$ (left to right). } 
\label{fig:E_sc}    
\end{figure}
In order to derive the effective spinon-chargon Hamiltonian \eqref{eq:H_eff_sc} from the mixD Hamiltonian \eqref{eq:mixD_Hamiltonian} for strong repulsive interactions $V, V- J_\perp\gg t_\parallel$ we perform a Schrieffer-Wolff transformation \cite{Schrieffer1966}, as schematically depicted in Fig. \ref{fig:2}c. Here, we adapt the notation introduced in Fig. \ref{fig:2}a with the spinon-chargon creation (annihilation) operators $\hat{f}_{i\mu\sigma}^{(\dagger)} $, i.e.
\begin{align}
    \hat{f}_{i\uparrow\sigma}^\dagger \ket{\dots \singlet \dots}=\ket{\dots \sigmahole\dots}
\end{align}
and 
\begin{align}
    \hat{f}_{i\uparrow\sigma} \ket{\dots \sigmahole \dots}=\ket{\dots \singlet \dots}\, .
\end{align}
In this notation, singlets are denoted by $\singlet$, spinon-chargon pairs by $\sigmahole$ or $\holesigma$ and chargon-chargon pairs by $\holes$. Hence, the sc vacuum, consisting of singlets on each rung of the ladder \cite{Bohrdt2021}, is denoted by $\ket{\dots \singlet \dots}$.

For the low-energy (Gutzwiller projected) sc hopping processes of sc's without neighbors we get 
\begin{align}
     \bra{\holesigma \singlet}\mathcal{H}_t\ket{\singlet \holesigma}= \frac{t_\parallel}{2}\left( \bra{\holesigma \updown } + \bra{\holesigma \downup} \right) \left( \ket{\holedown \upsigma } + \ket{\holeup \downsigma} \right)=\frac{t_\parallel}{2}\, .
\end{align}
Furthermore, second order processes without recombination to the chargon-chargon channel for isolated spinon-chargons without nearest neighbors have amplitude $-2\frac{t_\parallel^2}{J_\perp} \frac{3}{4}$, where the factor $2$ arises from the two directions in which the holes / particles can hop and the factor $\frac{3}{4}$ comes from the matrix element of this process. In addition, $\mathcal{H}_J$ gives a contribution $+J_\perp$ for every rung with a broken singlet w.r.t. the spinon-chargon vacuum (the ground state at half filling). Putting it all together, we arrive at the free spinon-chargon Hamiltonian
\begin{align}           
\mathcal{\hat{H}}^{sc,\mathrm{free}}_\mathrm{eff}&= \, \frac{t_\parallel}{2}\sum_{\langle ij\rangle} \sum_{\sigma,\mu} \mathcal{\hat{P}}_f\left(\hat{f}^\dagger_{j\mu\sigma}\hat{f}_{i,\mu\sigma}+\mathrm{h.c.}\right)\mathcal{\hat{P}}_f +\underbrace{\left(J_\perp+\frac{3}{2}\frac{t_\parallel^2}{(-J_\perp)}\right)}_ {=:\epsilon_0}\sum_{j\mu}\hat{n}^f_{i\mu}\, .
\label{eq:H_eff_sc_free}
\end{align}
As soon as two sc's occupy neighboring rungs, there is no contribution by second order processes without recombination to the chargon-chargon channel in the direction of the neighboring sc; since those terms are already included in the free Hamiltonian term \eqref{eq:H_eff_sc_free} a term 
\begin{align}           
-\frac{3}{2}\frac{t_\parallel^2}{(-J_\perp)}\sum_{\langle ij\rangle}\sum_{\mu\mu^\prime}\hat{n}^f_{i\mu}\hat{n}^f_{j\mu^\prime}
\end{align}
has to be added. Lastly, there are second order hopping terms via the high energy subspace $V$ (for recombination to the the triplet channel) and $V-J_\perp$ (singlet channel), schematically depicted in Fig. \ref{fig:2}c. These processes can be written in terms of chargon-chargon and spinon-chargon interactions, where chargon-chargon operators $\hat{b}_{i}^{(\dagger)}$ are defined by
\begin{align}
    \hat{b}_{i}^\dagger \ket{\dots \singlet \dots}=\ket{\dots \holes\dots}
\end{align}
and 
\begin{align}
    \hat{b}_{i} \ket{\dots \holes \dots}=\ket{\dots \singlet \dots}\, .
\end{align}
For the singlet channel we need to consider second order hopping processes from the sc to the cc channel and back, where the former is given by
\begin{align*}
    -\frac{t_\parallel}{\sqrt{2}}\sum_{\langle ij\rangle}\sum_{\mu\sigma} (-1)^\sigma\hat{f}_{i\sigma\mu}^\dagger\hat{f}_{j\bar{\sigma}\bar{\mu}}^\dagger\hat{b}_j
\end{align*}
and the latter analogously. The factor $(-1)^\sigma$ takes the sign structure of the singlets into account. The perturbative correction due to these processes is given by
\begin{align*}
    \Delta \mathcal{\hat{H}}_{ff} &= -\frac{2t_\parallel^2}{V-J_\perp}\frac{1}{2}\sum_{\langle ij\rangle}\sum_{\mu\mu^\prime}\sum_{\sigma\sigma^\prime} (-1)^\sigma (-1)^{\sigma^\prime} \hat{f}_{i\bar{\sigma}\bar{\mu}}^\dagger\hat{f}_{j\sigma\mu}^\dagger\hat{f}_{j\sigma^\prime\mu^\prime}  \hat{f}_{i\bar{\sigma}^\prime\bar{\mu}^\prime},
\end{align*}
which includes $\mu=\mu^\prime$ and $\mu=\bar{\mu}^\prime$ processes. Here $\mu=\bar{\mu}$ densotes the opposite leg of $\mu$, i.e. $\bar{0}=1$ and $\bar{1}=0$. The amplitude of the interaction, $-\frac{t_\parallel^2}{V-J_\perp}$, is attractive in the regime under consideration and diverges for $V\to J_\perp$. $\hat{\mathcal{H}}_{ff}$ can be represented using the operators $\vec{\hat{J}}$ defined in Eq. \eqref{eq:J_operators} and shown in Fig. \ref{fig:2}a, where $\mu=\mu^\prime$ processes correspond to $2\hat{J}_i^z\hat{J}_j^z-\frac{1}{2}$ and $\mu\neq \mu^\prime$ to $(\hat{J}_i^+\hat{J}_j^-+\hat{J}_i^-\hat{J}_j^+)$. By projecting onto the singlet channel using $\hat{P}_S:=- \hat{\vec{S}}_{i}\cdot \hat{\vec{S}}_{j}+\frac{1}{4}\hat{n}^f_{i}\hat{n}^f_{j}$, we arrive at
\begin{align*}
    \Delta \mathcal{\hat{H}}_{ff} &= -2\frac{t_\parallel^2}{V-J_\perp }\sum_{\langle ij\rangle}\left(
    \hat{J}_i^+\hat{J}_j^-+\hat{J}_i^-\hat{J}_j^+-2\hat{J}_i^z\hat{J}_j^z+\frac{1}{2}
    \right)\left( -\hat{\vec{S}}_{i}\cdot \hat{\vec{S}}_{j}+\frac{1}{4}\hat{n}^f_{i}\hat{n}^f_{j}\right)\notag \\
    &= -4\frac{t_\parallel^2}{V-J_\perp }\sum_{\langle ij\rangle}\left(-
    \vec{\hat{J}}_i\cdot \vec{\hat{J}}_j+\frac{1}{4}
    \right)\left( -\hat{\vec{S}}_{i}\cdot \hat{\vec{S}}_{j}+\frac{1}{4}\hat{n}^f_{i}\hat{n}^f_{j}\right)\, .
\end{align*}
For the triplet channel we get
\begin{align*}
    \Delta \mathcal{\hat{H}}_{ff} &= -2\frac{t_\parallel^2}{V}\frac{1}{2}\sum_{\langle ij\rangle}\sum_{\mu\mu^\prime}\sum_{\sigma\sigma^\prime} \hat{f}_{i\bar{\sigma}\bar{\mu}}^\dagger\hat{f}_{j\sigma\mu}^\dagger\hat{f}_{j\sigma^\prime\mu^\prime}  \hat{f}_{i\bar{\sigma}^\prime\bar{\mu}^\prime}\notag \\
    &=-2\frac{t_\parallel^2}{V }\sum_{\langle ij\rangle}\left(
    J_i^+J_j^-+J_i^-J_j^+-2J_i^zJ_j^z+\frac{1}{2}
    \right)\left( \hat{\vec{S}}_{i}\cdot \hat{\vec{S}}_{j}+\frac{3}{4}\hat{n}^f_{i}\hat{n}^f_{j}\right)\notag \\
    &=-4\frac{t_\parallel^2}{V }\sum_{\langle ij\rangle}\left(-
    \vec{\hat{J}}_i\cdot \vec{\hat{J}}_j+\frac{1}{4}
    \right)\left( \hat{\vec{S}}_{i}\cdot \hat{\vec{S}}_{j}+\frac{3}{4}\hat{n}^f_{i}\hat{n}^f_{j}\right)\, ,
\end{align*}
where $\hat{P}_T:=\left( \hat{\vec{S}}_{i}\cdot \hat{\vec{S}}_{j}+\frac{3}{4}\hat{n}^f_{i}\hat{n}^f_{j}\right)$ projects onto the triplet channel.\\

The resulting effective Hamiltonian Eq. \eqref{eq:H_eff_sc} features both repulsive and attractive interactions. Since the attraction is mediated by recombinations into the cc channel, its amplitude is proportional to $-\frac{t_\parallel^2}{V-J_\perp}$, yielding dominant attraction and finite binding energies of the effective model if $V\approx V_c= J_\perp$ for small $t_\parallel/J_\perp $, see Fig. \ref{fig:E_sc} left. For larger $t_\parallel/J_\perp $, binding is stabilized and hence the resonance shifts to higher $V_c>J_\perp$. This is also observed in numerical simulations of the effective model \eqref{eq:H_eff_sc}, where a doping regime with finite positive binding energies emerges for $V=5J_\perp$ and $t_\parallel/J_\perp = 1.0, 3.0$, see Fig. \ref{fig:E_sc} right. Note that the sc description can only be applied up to $\delta=50\%$, corresponding to full sc filling.

\subsection{The bond-ordered density wave at $\delta=0.5$ \label{appendix:BOWD}}

For $\delta=0.5$ and $V, J_\perp\gg t_\parallel$ the effective sc Hamiltonian \eqref{eq:H_eff_sc} becomes 
\begin{align}
    \hat{H}_\mathrm{eff} =& -4\frac{t_\parallel^2}{V-J_\perp }\sum_{j}\left(-\vec{\hat{J}}_{j+1}\cdot \vec{\hat{J}}_j+\frac{1}{4}
\right)\left(- \vec{S}_{j+1}\cdot \vec{S}_{j}+\frac{1}{4}\hat{n}_{j+1}\hat{n}^f_{j}\right)\notag\\
&-4\frac{t_\parallel^2}{V }\sum_{j}\left(
-\vec{\hat{J}}_{j+1}\cdot \vec{\hat{J}}_j+\frac{1}{4}
\right)\left( \vec{S}_{j+1}\cdot \vec{S}_{j}+\frac{3}{4}\hat{n}^f_{j+1}\hat{n}^f_{j}\right),
\label{eq:sc_Ham_half_filling}
\end{align}
since in the case of half-filling (i.e. maximal sc filling) the first term of Eq. \eqref{eq:H_eff_sc} vanishes due to the Gutzwiller projection $\mathcal{\hat{P}}_f$. Furthermore, the second and third term as well as the $-\frac{1}{4}$ and $+\frac{3}{4}$ terms in the singlet and triplet projectors give constant contributions in this case. 

Note that for $J_\perp=0$ singlets and triplets are degenerate:
\begin{align}
    \hat{H}_\mathrm{eff}^{J_\perp=0} =& -4\frac{t_\parallel^2}{V}\sum_{j}\left(-\vec{\hat{J}}_{j+1}\cdot \vec{\hat{J}}_j+\frac{1}{4}\right).
    \label{eq:Heff_sc_J=0}
\end{align}
For small $J_\perp$ we can Taylor expand
\begin{align}
    \frac{t_\parallel^2}{V-J_\perp } = \frac{t_\parallel^2}{V}\left(\frac{1}{1-\frac{J_\perp}{V}} \right)\approx \frac{t_\parallel^2}{V}\left( 1+\frac{J_\perp}{V}\right)+\dots\,.
\end{align}
In this case Eq. \eqref{eq:sc_Ham_half_filling} becomes
\begin{align}
    \hat{H}_\mathrm{eff} = &   \hat{H}_\mathrm{eff}^{J_\perp=0}-4\frac{t_\parallel^2 J_\perp}{V^2}\sum_{j}\left(-\vec{J}_{j+1}\cdot \vec{J}_{j}+\frac{1}{4}
\right)\left(-\vec{S}_{j+1}\cdot \vec{S}_{j}+\frac{1}{4}\right)\notag \\
= & \hat{H}_\mathrm{eff}^{J_\perp=0}-4\frac{t_\parallel^2 J_\perp}{V^2}\sum_{j}(\vec{J}_{j+1}\cdot \vec{J}_{j})(\vec{S}_{j+1}\cdot \vec{S}_{j})
+\frac{t_\parallel^2 J_\perp}{V^2}\sum_{j}\vec{J}_{j+1}\cdot \vec{J}_{j}
+\frac{t_\parallel^2 J_\perp}{V^2}\sum_{j}\vec{S}_{j+1}\cdot \vec{S}_{j}
+\mathrm{const.}\,.
\label{eq:HBvsVBS}
\end{align}
This Hamiltonian includes competing terms that favor either Heisenberg (HB) order (first, 3rd and 4th term) or alternating singlet/triplet order (second term). The second term can have lower energies for a valence-bond crystal (VBS) state of spin and isospins, i.e. an alternating pattern of singlets (no singlets) on bonds $\langle 2j,2j+1 \rangle $ ($\langle 2j+1,2j+2 \rangle $) as illustrated in Fig. \ref{fig:4}a. This can be seen from comparing variational energies:
\begin{itemize}
    \item The energy per bond of a Heisenberg AFM is $\left(\frac{1}{4}-\mathrm{ln}(2)\right)$. Consequently, the energy per bond from Eq. \eqref{eq:HBvsVBS} for HB order in both spin and isospin is $$(E_0^\mathrm{HB})^2 =-(\frac{1}{4}-\mathrm{ln}(2))^2\approx -(0.443)^2=-0.196$$
(squared because for spins and legs each).
    \item The VBS state consists of leg and spin singlets on bonds $\langle 2j,2j+1 \rangle $, i.e. on $50\%$ of the bonds. Since each leg and spin singlets contribute energy $\frac{3}{4}$ per bond, we have 
$$E_0=-\frac{1}{2}\left( \frac{3}{4}\right)^2 \approx -0.28$$. 
\end{itemize}
This comparison shows that the second term in Eq. \eqref{eq:HBvsVBS} favors a VBS state. Implications on expectation values like $\langle \vec{\hat{J}}_i\cdot  \vec{\hat{J}}_{i+1}\rangle$ and $\langle \vec{\hat{S}}_i\cdot  \vec{\hat{S}}_{i+1}\rangle$ and the additional Friedel oscillation peaks, associated with excitations from the BODW order, are discussed in the main text and below.

\end{document}